\documentclass[a4paper,onecolumn,11pt,unpublished]{quantumarticle}
\pdfoutput=1

\usepackage[T1]{fontenc}
\usepackage[utf8]{inputenc}

\usepackage[english]{babel}
\usepackage{csquotes}

\usepackage{amsmath}
\usepackage{hyperref}
\usepackage{cleveref}

\usepackage[numbers,sort&compress]{natbib}

\usepackage{url}						

\usepackage{subcaption} 
\usepackage{graphicx}
\usepackage{color}                    				
\usepackage{tikz}
\usepackage{tikz-3dplot}
\usetikzlibrary{positioning,intersections,shadings,arrows,shapes,snakes,automata,backgrounds,petri,topaths,matrix}

\usepackage{xstring}
\usepackage{pgfkeys}
\usepackage{xparse}

\usepackage{stmaryrd}					
\usepackage{amssymb}					
\usepackage{amsthm}					
\usepackage{bm} 

\theoremstyle{plain}
\newtheorem{thm}{Theorem}
\crefname{thm}{theorem}{theorems}
\Crefname{thm}{Theorem}{Theorems}
\theoremstyle{plain}

\crefname{prop}{proposition}{propositions}
\Crefname{prop}{Proposition}{Propositions}
\theoremstyle{plain}
\theoremstyle{definition}
\newtheorem{defn}{Definition}
\crefname{defn}{definition}{definitions}
\Crefname{defn}{Definition}{Definitions}
\theoremstyle{plain}
\newtheorem{lem}{Lemma}
\crefname{lem}{lemma}{lemmas}
\Crefname{lem}{Lemma}{Lemmas}
\theoremstyle{plain}

\crefname{cor}{corollary}{corollaries}
\Crefname{cor}{Corollary}{Corollaries}
\theoremstyle{plain}

\crefname{rem}{remark}{remarks}
\Crefname{rem}{Remark}{Remarks}

\begin{document}
\newcommand{\covtrip}{covariance triple} 
\newcommand{\coma}[1]{{\color{red}{#1}}}

\renewcommand{\labelitemi}{$\bullet$}

\newcommand{\indep}{\rotatebox[origin=c]{90}{$\models$}}
\newcommand{\ot}{\otimes}
\newcommand{\eps}{\epsilon}
\newcommand{\sth}{ \ \mathrm{s.t.} \ }
\newcommand{\trans}{\mathrm{T}}
\newcommand{\partr}[2]{\operatorname{tr}_{#2}\left( #1 \right)}
\newcommand{\trdist}[2]{\operatorname{D}\left( #1 | #2 \right)}
\newcommand{\lin}[1]{\mathcal{L}\left( #1 \right)} 
\newcommand{\linH}[1]{\mathcal{L}_\mathrm{H}\left( #1 \right)}

\newcommand{\epi}{\operatorname{epi}}

\newcommand{\ket}[1]{| #1 \rangle}
\newcommand{\bra}[1]{\langle #1 |}
\newcommand{\braket}[2]{\langle #1|#2\rangle}
\newcommand{\ketbra}[2]{|#1\rangle\!\langle#2|}
\newcommand{\proj}[1]{\ketbra{#1}{#1}}
\newcommand{\identity}{\openone}
\newcommand{\id}{\mathbbm{1}}
\newcommand{\cptp}{\mathcal{E}}

\newcommand{\acal}{\mathcal{A}}
\newcommand{\bcal}{\mathcal{B}}
\newcommand{\ccal}{\mathcal{C}}
\newcommand{\dcal}{\mathcal{D}}
\newcommand{\ecal}{\mathcal{E}}
\newcommand{\fcal}{\mathcal{F}}
\newcommand{\lcal}{\mathcal{L}}
\newcommand{\ocal}{\mathcal{O}}
\newcommand{\vcal}{\mathcal{V}}
\newcommand{\zcal}{\mathcal{Z}}
\newcommand{\kcal}{\mathcal{K}}
\newcommand{\hcal}{\mathcal{H}}
\newcommand{\tcal}{\mathcal{T}}
\newcommand{\scal}{\mathcal{S}}
\newcommand{\pcal}{\mathcal{P}}
\newcommand{\ucal}{\mathcal{U}}

\newcommand{\sbuild}[2]{\left\{#1\,\middle|\,#2\right\}}

\newcommand{\supp}[1]{\operatorname{supp}\left({#1}\right)}

\NewDocumentCommand{\bops}{ o o }
{
  \lcal\IfValueT{#1}{\!\left({#1}\IfValueT{#2}{,{#2}}\right)}
}

\NewDocumentCommand{\saops}{ o }
{
  \lcal_{s}\IfValueT{#1}{\!\left({#1}\right)}
}

\NewDocumentCommand{\posops}{ o }
{
  \lcal_{s}^{+}\IfValueT{#1}{\!\left({#1}\right)}
}

\NewDocumentCommand{\trops}{ o }
{
  \tcal\IfValueT{#1}{\!\left({#1}\right)}
}

\NewDocumentCommand{\dops}{ o }
{
  \scal\IfValueT{#1}{\!\left({#1}\right)}
}

\NewDocumentCommand{\probs}{ m m }
{
  \pcal\!\left({#1}, {#2}\right)
}

\NewDocumentCommand{\seq}{ m o o o }
{
  \left({#1}\right)_{ \IfValueT{#2}{#2\IfValueT{#3}{={#3}} }  }^{\IfValueT{#4}{#4}}
}

\NewDocumentCommand{\cov}{ O{} O{} o }
{
  \mathcal{C}_{{#1}{#2}}\IfValueT{#3}{\!\left[{#3}\right]}
}

\NewDocumentCommand{\tr}{ m o }
{
  \operatorname{tr}{\IfValueT{#2}{_{#2}}}\left( #1 \right)
}

\newcommand{\vspan}[1]{\operatorname{span}\left(#1\right)}

\newcommand{\fmath}{\mathbb{F}}
\newcommand{\smath}{\mathbb{S}}
\newcommand{\abs}[1]{{\left\lvert{#1}\right\rvert}}
\newcommand{\then}{\Rightarrow}
\newcommand\Eb{\mathbb{E}}
\newcommand\N{\mathbb{N}}
\newcommand\Z{\mathbb{Z}}
\newcommand\Q{\mathbb{Q}}
\newcommand\ps{{(\Omega, \mathcal{F}, \mathbb{P})}}
\newcommand{\R}{\mathbb{R}}
\newcommand{\B}{\mathbb{B}}
\newcommand{\K}{\mathbb{K}}
\newcommand\Pb{\mathbb{P}}
\newcommand{\exR}{\overline{\R}}
\newcommand{\Cx}{\mathbb{C}}
\newcommand{\T}{\mathbb{T}}
\newcommand\s{$\sigma$\textrm{-field} }
\newcommand\notsubset{\subset\hspace{-3.5mm}{/}\hspace{1.5mm}}

\def\del{\dot{\Delta}}
\def\lap{\nabla}
\def\invf{\mathcal{F}^{-1}}
\def\dj{\delta_j}
\def\apo{(I + \delta B)^{-1}}
\def\T{\mathbb{T}}
\def\A{\mathrm{A}}
\def\mcal{\mathcal{M}}
\def\divv{\mathrm{div}}
\def\rE{\mathrm{E}}
\def\e{\mathbb{E}}


\newcommand{\born}{\bm{r}_0}

\newcommand{\boap}{\bm{\overline{a}}}
\newcommand{\vproj}[2]{\pi_{{#1}{#2}}}
\newcommand{\boa}{\bm{a}}
\newcommand{\bob}{\bm{b}}
\newcommand{\boc}{\bm{c}}
\newcommand{\bod}{\bm{d}}
\newcommand{\boe}{\bm{e}}
\newcommand{\bof}{\bm{f}}
\newcommand{\bog}{\bm{g}}
\newcommand{\boh}{\bm{h}}
\newcommand{\boj}{\bm{j}}
\newcommand{\bok}{\bm{k}}
\newcommand{\bol}{\bm{l}}
\newcommand{\bom}{\bm{m}}
\newcommand{\bon}{\bm{n}}
\newcommand{\boo}{\bm{o}}
\newcommand{\bop}{\bm{p}}
\newcommand{\boq}{\bm{q}}
\newcommand{\bor}{\bm{r}}
\newcommand{\bos}{\bm{s}}
\newcommand{\bou}{\bm{u}}
\newcommand{\bov}{\bm{v}}
\newcommand{\bow}{\bm{w}}
\newcommand{\boex}{\bm{x}}
\newcommand{\boy}{\bm{y}}
\newcommand{\boz}{\bm{z}}

\newcommand{\bogam}{\bm{\gamma}}
\newcommand{\boalpha}{\bm{\alpha}}

\newcommand{\bosig}{{\boldsymbol\sigma}}

\newcommand{\opa}{\operatorname{A}}
\newcommand{\opb}{\operatorname{B}}
\newcommand{\opc}{\operatorname{C}}
\newcommand{\opd}{\operatorname{D}}
\newcommand{\ope}{\operatorname{E}}
\newcommand{\opf}{\operatorname{F}}
\newcommand{\opg}{\operatorname{G}}
\newcommand{\oph}{\operatorname{H}}
\newcommand{\opi}{\operatorname{I}}
\newcommand{\opj}{\operatorname{J}}
\newcommand{\opk}{\operatorname{K}}
\newcommand{\opl}{\operatorname{L}}
\newcommand{\opm}{\operatorname{M}}
\newcommand{\opn}{\operatorname{N}}
\newcommand{\opo}{\operatorname{O}}
\newcommand{\opp}{\operatorname{P}}
\newcommand{\opq}{\operatorname{Q}}
\newcommand{\opr}{\operatorname{R}}
\newcommand{\ops}{\operatorname{S}}
\newcommand{\opt}{\operatorname{T}}
\newcommand{\opu}{\operatorname{U}}
\newcommand{\opv}{\operatorname{V}}
\newcommand{\opw}{\operatorname{W}}
\newcommand{\opx}{\operatorname{X}}
\newcommand{\opy}{\operatorname{Y}}
\newcommand{\opz}{\operatorname{Z}}

\newcommand{\oppi}{\operatorname{\pi}}

\newcommand{\defeq}{:=}

\newcommand{\expr}[1]{\left\langle {#1}\right\rangle_\rho} 
\newcommand{\expnr}[1]{\left\langle {#1}\right\rangle} 

\DeclareDocumentCommand{\sdev}{ O{} m }{
  {\Delta_{{#1}} {#2}}
}
\DeclareDocumentCommand{\var}{ O{} m }{
  {\Delta^2_{#1} {#2}}
}

\DeclareDocumentCommand{\qbit}{ O{} O{+} O{\frac{1}{2}} m }{
  {{#3}\left({#1}\opi\, {#2}\, {#4}\cdot\bosig\right)}
}

\newcommand{\varr}[1]{{\var[\rho]{{#1}}}}
\newcommand{\sdevr}[1]{{\sdev[\rho]{{#1}}}}

\newcommand{\sdevmin}[1]{{\sdev{#1}_\text{min}}}
\newcommand{\varmin}[1]{{\var{#1}_\text{min}}}

\newcommand{\sdevmax}[1]{{\sdev{#1}_\text{max}}}
\newcommand{\varmax}[1]{{\var{#1}_\text{max}}}

\newcommand{\ft}[1]{\widetilde{#1}} 
\newcommand{\Cos}[1]{\cos\left(#1\right)} 
\newcommand{\Sin}[1]{\sin\left(#1\right)} 

\newcommand{\bracks}[1]{\left(#1\right)} 
\newcommand{\sbracks}[1]{\left[#1\right]} 
\newcommand{\com}[2]{\left[#1,#2\right]} 
\newcommand{\comm}[2]{#1 #2 - #2 #1} 
\newcommand{\acom}[2]{#1 #2 + #2 #1} 

\newcommand{\si}{\mathcal{S}}
\newcommand{\hi}{\mathcal{H}}

\renewcommand{\Re}{\operatorname{Re}}
\renewcommand{\Im}{\operatorname{Im}}
\newcommand{\ival}{I}
\newcommand{\PUR}[3]{\operatorname{PUR}_{#1}\left({#2},{#3}\right)}

\pgfkeys{
  /drawSemiCircle/.is family, /drawSemiCircle,
  defaults/.style = {scale = \textwidth,
    aAngle = 0,
    bAngle = 0,
    rAngle = 0,
    rLength = 1,
    drawBPrime = false,
    drawRightAngles = true,
    rightAngleScale = 0.07},
  scale/.estore in = \scale,
  aAngle/.estore in = \aAngle,
  bAngle/.estore in =\bAngle,
  rAngle/.estore in =\rAngle,
  rLength/.estore in =\rLength,
  drawBPrime/.estore in =\drawBPrime,
  rSub/.estore in =\rSub,
  drawRightAngles/.estore in =\drawRightAngles,
  rightAngleScale/.estore in =\rightAngleScale,
}

\newcommand{\DrawSemiCircle[2]}{%
  \pgfkeys{/drawSemiCircle, defaults, #1}%
  \begin{tikzpicture}[scale=\scale/2cm,>=stealth] 
    \draw (-1,0) -- (1,0);

    \clip (-1cm, 0cm) rectangle (1.2cm, 1.2cm);

    \draw (0cm,0cm) circle (1cm);

    \draw [name=vect,thick,->] (0,0) -- node[pos=1,fill=none,label=\aAngle:$\boa$] {} (\aAngle:1cm);
    \draw [name=vect,thick,->] (0,0) -- node[pos=1,fill=none,label=\bAngle:$\bob$] {} (\bAngle:1cm);
    \IfEqCase{\drawBPrime}{
      {true}{
        \pgfmathsetmacro{\bPrimeAngle}{\bAngle-90};
        \draw [name=vect,thick,->] (0,0) -- node[pos=1, left  = 0.1 ,fill=none] {$\bob^\prime$} (\bPrimeAngle:1cm);
        \IfEqCase{\drawRightAngles}{
          {true}{
            \pgfmathsetmacro{\midAngle}{(\bPrimeAngle+\bAngle)/2}
            \pgfmathsetmacro{\midLength}{sqrt(2)*\rightAngleScale}
            \draw [name=rangle] (\bAngle:\rightAngleScale) -- (\midAngle:\midLength) -- (\bPrimeAngle:\rightAngleScale);
          }
          {false}{}
        }
      }
      {false}{}
    }
    \draw [name=vect,thick,->] (0,0) -- node[midway, fill=white] {$\bor_{\rSub}$} (\rAngle:\rLength);

    \pgfmathsetmacro{\ra}{1*\rLength*cos(\aAngle-\rAngle)}
    \pgfmathsetmacro{\rb}{1*\rLength*cos(\bAngle-\rAngle)}

    \draw [name=vect,thick,->] (0,0) -\- node[pos=1, below right = 0.06 and 0.03 ,fill=none] {$\boa^*$} (\aAngle:\ra);
    \draw [name=vect,thick,->] (\aAngle:\ra) -- node[midway,fill=white] {$\bm{x}$} (\rAngle:\rLength);
    \IfEq{\rb}{0}{
      \IfEqCase{\drawRightAngles}{
        {true}{
          \pgfmathsetmacro{\midAngle}{(\rAngle+\bAngle)/2}
          \pgfmathsetmacro{\midLength}{sqrt(2)*\rightAngleScale}
          \draw [name=rangle] (\rAngle:\rightAngleScale) -- (\midAngle:\midLength) -- (\bAngle:\rightAngleScale);
        }
        {false}{}
      }
    }
    {\draw [name=vect,thick,->] (0,0) -- node[pos=1, below left  = 0.05 and 0.03 ,fill=none] {$\bob^*$} (\bAngle:\rb);
      \draw [name=vect,thick,->] (\bAngle:\rb) -- node[midway,fill=white] {$\bm{y}$} (\rAngle:\rLength);}

    \IfEq{\ra}{0}{}{
      \IfEqCase{\drawRightAngles}{
        {true}{
          \pgfmathsetmacro{\rx}{\rLength*cos(\rAngle)}
          \pgfmathsetmacro{\ry}{\rLength*sin(\rAngle)}

          \pgfmathsetmacro{\rax}{\ra*cos(\aAngle)}
          \pgfmathsetmacro{\ray}{\ra*sin(\aAngle)}
          \pgfmathsetmacro{\xxUnNormed}{\rx - \rax}
          \pgfmathsetmacro{\xyUnNormed}{\ry - \ray}
          \pgfmathsetmacro{\xx}{\xxUnNormed / sqrt(\xxUnNormed*\xxUnNormed + \xyUnNormed*\xyUnNormed)}
          \pgfmathsetmacro{\xy}{\xyUnNormed / sqrt(\xxUnNormed*\xxUnNormed + \xyUnNormed*\xyUnNormed)}
          \pgfmathsetmacro{\aStartX}{\rax * (1 + \rightAngleScale/\ra)}
          \pgfmathsetmacro{\aStartY}{\ray * (1 + \rightAngleScale/\ra)}
          \pgfmathsetmacro{\aMidX}{\aStartX  + \rightAngleScale*\xx}
          \pgfmathsetmacro{\aMidY}{\aStartY  + \rightAngleScale*\xy}
          \pgfmathsetmacro{\aEndX}{\rax + \rightAngleScale*\xx}
          \pgfmathsetmacro{\aEndY}{\ray + \rightAngleScale*\xy}
          \draw [name=rangle] (\aStartX,\aStartY) --  (\aMidX,\aMidY) -- (\aEndX,\aEndY);
        }
        {false}{}
      }
    }
    \IfEq{\rb}{0}{}{
      \IfEqCase{\drawRightAngles}{
        {true}{
          \pgfmathsetmacro{\rbx}{\rb*cos(\bAngle)}
          \pgfmathsetmacro{\rby}{\rb*sin(\bAngle)}
          \pgfmathsetmacro{\yxUnNormed}{\rx - \rbx}
          \pgfmathsetmacro{\yyUnNormed}{\ry - \rby}
          \pgfmathsetmacro{\yx}{\yxUnNormed / sqrt(\yxUnNormed*\yxUnNormed + \yyUnNormed*\yyUnNormed)}
          \pgfmathsetmacro{\yy}{\yyUnNormed / sqrt(\yxUnNormed*\yxUnNormed + \yyUnNormed*\yyUnNormed)}
          \pgfmathsetmacro{\bStartX}{\rbx * (1 + \rightAngleScale/\rb)}
          \pgfmathsetmacro{\bStartY}{\rby * (1 + \rightAngleScale/\rb)}
          \pgfmathsetmacro{\bMidX}{\bStartX  + \rightAngleScale*\yx}
          \pgfmathsetmacro{\bMidY}{\bStartY  + \rightAngleScale*\yy}
          \pgfmathsetmacro{\bEndX}{\rbx + \rightAngleScale*\yx}
          \pgfmathsetmacro{\bEndY}{\rby + \rightAngleScale*\yy}
          \draw [name=rangle] (\bStartX,\bStartY) --  (\bMidX,\bMidY) -- (\bEndX,\bEndY);
        }
        {false}{}
      }
    }
  \end{tikzpicture}
}
\DeclareDocumentCommand{\DrawSphere}{ m m m }{%
  \begin{tikzpicture}
    \tdplotsetmaincoords{10}{0}
    \tdplotsetrotatedcoords{90}{90}{-90}
    \coordinate (O) at (0,0,0);
    \pgfmathsetmacro{\scale}{0.1cm}

    \pgfmathsetmacro{\aAngle}{{#1}}
    \pgfmathsetmacro{\bAngle}{{#2}}
    \pgfmathsetmacro{\rZeroAngle}{{#3}}
    \pgfmathsetmacro{\rOneAngle}{\aAngle - abs(\aAngle - \rZeroAngle)}
    \pgfmathsetmacro{\rOneLength}{\rOneAngle < 0 ? sqrt( 1 + (cos(\aAngle - \bAngle) / cos(\bAngle - \aAngle + 90))^2  ) *  abs(cos(\rZeroAngle - \aAngle)) : 1}
    \pgfmathsetmacro{\rOneAngle}{\rOneAngle < 0 ? 0 : \rOneAngle}

    \pgfmathsetmacro{\bPrimeAngle}{\bAngle - 90}

    \tdplotsetcoord{a}{\scale}{90}{\aAngle}
    \tdplotsetcoord{b}{\scale}{90}{\bAngle}
    \tdplotsetcoord{bPrime}{\scale}{90}{\bPrimeAngle}
    \tdplotsetcoord{rZero}{\scale}{90}{\rZeroAngle}
    \tdplotsetcoord{rOne}{\rOneLength*\scale}{90}{\rOneAngle}

    \tdplotdrawarc[tdplot_screen_coords]{(O)}{\scale}{0}{360}{}{}

    \draw [thick,->] (O) -- node[pos=1,fill=none, label={\aAngle:$\boa$}] {} (a);
    \IfEq{\bAngle}{\rZeroAngle}{
      \draw [thick,->] (O) -- node[pos=1, fill=none, label={\bAngle:$\bob = \bor_0$}] {} (b);
    }{
      \draw [thick,->] (O) -- node[pos=1, fill=none, label={\bAngle:$\bob$}] {} (b);
      \draw [thick,->] (O) -- node[pos=1, fill=none, label={\rZeroAngle:$\bor_0$}] {} (rZero);
    }

    \IfEq{\bPrimeAngle}{\rOneAngle}{
      \IfEq{\rOneLength}{1}{
        \draw [thick,->] (O) -- node[pos=1, fill=none, label={\bPrimeAngle:$\bob^\prime = \bor_1$}] {} (bPrime);
      }{
        \draw [thick,->] (O) -- node[pos=1, fill=none, label={\bPrimeAngle:$\bob^\prime=\hat\bor_1$}] {} (bPrime);
        \draw [thick,->] (O) -- node[pos=1, fill=none, label={[label distance=0.2*\scale]120:$\bor_1$}] {} (rOne);
      }
    }{
      \draw [thick,->] (O) -- node[pos=1, fill=none, label={\bPrimeAngle:$\bob^\prime$}] {} (bPrime);
      \draw [thick,->] (O) -- node[pos=1, fill=none, label={\rOneAngle:$\bor_1$}] {} (rOne);
    }
    \pgfmathsetmacro{\bRZeroCircleRad}{abs(sin(\bAngle - \rZeroAngle))}
    \pgfmathsetmacro{\bRZeroCircleLoc}{abs(cos(\bAngle - \rZeroAngle))}
    \tdplotdrawarc[tdplot_rotated_coords]{(0,0,\bRZeroCircleLoc*\scale)}{\bRZeroCircleRad*\scale}{0}{180}{anchor=north}{  }
    \tdplotdrawarc[tdplot_rotated_coords, style=dashed]{(0,0,\bRZeroCircleLoc*\scale)}{\bRZeroCircleRad*\scale}{180}{360}{anchor=north}{  }

    \pgfmathsetmacro{\bROneCircleRad}{abs(sin(\bAngle - \rOneAngle))}
    \pgfmathsetmacro{\bROneCircleLoc}{cos(\bAngle - \rOneAngle)}
    \tdplotdrawarc[tdplot_rotated_coords]{(0,0,\bROneCircleLoc*\scale)}{\bROneCircleRad*\scale}{0}{180}{anchor=north}{  }
    \tdplotdrawarc[tdplot_rotated_coords, style=dashed]{(0,0,\bROneCircleLoc*\scale)}{\bROneCircleRad*\scale}{180}{360}{anchor=north}{  }

    \tdplotsetrotatedcoords{\aAngle}{90}{-90}
    \pgfmathsetmacro{\aRCircleRad}{abs(sin(\aAngle - \rZeroAngle))}
    \pgfmathsetmacro{\aRCircleLoc}{abs(cos(\aAngle - \rZeroAngle))}
    \tdplotdrawarc[tdplot_rotated_coords]{(0,0,\aRCircleLoc*\scale)}{\aRCircleRad*\scale}{0}{180}{anchor=north}{  }
    \tdplotdrawarc[tdplot_rotated_coords, style=dashed]{(0,0,\aRCircleLoc*\scale)}{\aRCircleRad*\scale}{180}{360}{anchor=north}{  }
  \end{tikzpicture}
}

\newcommand{\grp}[3]{g\left({#1}, {#2}, {#3}\right)}
\newcommand{\map}[3]{R_{{#1}, {#2}, {#3}}}

\newcommand{\adj}[1]{{\operatorname{adj}\left({#1}\right)}}

\DeclareDocumentCommand{\cyc}{ O{n} }{
  {\mathbb{Z}_n}
}

\newcommand{\norm}[1]{{\left\lVert{#1}\right\rVert}}

\DeclareDocumentCommand{\pnorm}{ m O{p} }{
  \norm{{#1}}_{#2}
}
\newcommand{\fdiv}[3]{D_{#1}\left({#2}\parallel{#3}\right)}

\newcommand{\ovms}{\mathcal{M}}


\newcommand{\MOD}[1]{#1_{\mathrm{mod}}} 
\newcommand{\psdev}[1]{\tilde\Delta{\big(#1\big)}} 
\DeclareDocumentCommand{\ws}{ m o }{
  \operatorname{W}\IfValueT{ #2 }{_{#2}}{ } \left({#1}\right)
}

\DeclareDocumentCommand{\wsDev}{ m o }{
  \Delta\IfValueT{ #2 }{_{#2}}{ } \left({#1}\right)
}

\DeclareDocumentCommand{\lPVar}{ m m O{p} }{
  \delta_{#3}\left({#1}, {#2}\right)
}

\newcommand{\dom}[1]{\mathcal{D}(#1)}
\newcommand{\E}{\mathsf{E}} 
\newcommand{\iunit}{\mathrm{i}} 
\newcommand{\IT}{{I}_T} 
\newcommand{\IK}{{I}_K} 
\newcommand{\Ewp}{\mathsf E}
\newcommand{\Efc}{\mathsf F}

\newcommand{\hh}{^H} 
\renewcommand{\ss}{^S} 

\newcommand{\expe}[2]{\left\langle{#1}\right\rangle_{#2}}

\DeclareDocumentCommand{\borel}{ m o }{
  \mathcal{B}\left({#1}\IfValueT{#2}{,{#2}} \right)
}

\newcommand{\width}[2]{W_{#2}\left({#1}\right)}

\DeclareDocumentCommand{\calibError}{ m m m o }{
  \Delta^{\IfValueTF{#4}{#4}{c}}_{#3}\left({#1}, {#2}\right)
}

\newcommand{\spec}[1]{\sigma\left({#1}\right)}
\newcommand{\res}[1]{\rho\left({#1}\right)}

\title{Simplifying measurement uncertainty with quantum symmetries}

\author{Oliver Reardon-Smith}
\affiliation{Department of mathematics, University of York, York, UK}
\orcid{0000-0002-0124-1389}

\maketitle
\begin{abstract}
      Determining the measurement uncertainty region is a difficult problem for generic sets of observables. For this reason the literature on exact measurement uncertainty regions is focused on symmetric sets of observables, where the symmetries are used to simplify the calculation. We provide a framework to systematically exploit available symmetries, formulated in terms of covariance, to simplify problems of measurement uncertainty. Our key result is that for a wide range figures of merit the optimal compatible approximations of covariant target observables are themselves covariant. This substantially simplifies the problem of determining measurement uncertainty regions for cases where it applies, since the space of covariant observables is typically much smaller than that of all observables.
An intermediate result, which may be applicable more broadly, is the existence and characterisation of a covariantisation map, mapping observables to covariant observables. Our formulation is applicable to finite outcome observables on separable Hilbert spaces. We conjecture that the restriction of finite outcomes may be lifted, and explore some of the features a generalisation must have.
We demonstrate the theorem by deriving measurement uncertainty regions for three mutually orthogonal Pauli observables, and for phase space observables in arbitrary finite dimensions.
\end{abstract}

\section{Introduction}

The idea of defining an error for quantum measurements based on differences in their statistics was proposed as early as 1988 by Ludwig~\cite{ludwig-foundations-of-qm}, however the approach to measurement uncertainty we take here is based on a series of papers of Busch, Lahti and Werner~\cite{6773660Werner:2004:URJ:2011593.2011606,PhysRevLett.111.160405,blw-meas-uncertainty}. Those authors developed measurement uncertainties in terms of the Wasserstein distance, the total variation norm, and Monge metric and applied these ideas to the phase space observables of a particle free to move in one dimension, and those for a particle restricted to the vertices of a regular polygon. The case of the phase space of a particle on a ring was investigated by Busch Kiukas and Werner~\cite{sharp-ur-num-angle}, while Werner studied general phase spaces in ref.~\cite{Werner2016}. Non phase-space case studies have also been studied, including pairs of qubit observables~\cite{BuschHeinosaari2008,BuschLahtiWerner2014,BuschBullock2018} and angular momentum observables~\cite{DammeierSchwonnekWerner2015}. 

A key feature shared by these examples is symmetry, exhibited by the existence of a system of covariance, which is used to simplify the calculations. The study of measurement uncertainty in the absence of simplifying symmetries has been more restricted, however algorithms to compute the uncertainty region are known for several cases where the problem may be reduced to a semidefinite program~\cite{srw-meas-uncertainty-finite}. 

We do not seek to remedy this situation, but instead provide a systematic way of exploiting available symmetries to simplify problems of measurement uncertainty. We define a ``covariantisation'' map which, given a fixed system of covariance, maps finite outcome observables to covariant ones. We employ this to provide conditions defining a class of error measures, for which we prove that replacing an approximating observable with a covariant one does not increase the error. The idea of a covariantisation map is not a new one, a method based on an invariant mean was defined by Werner in~\cite{6773660Werner:2004:URJ:2011593.2011606}. However construction is suitable only for phase space observables, and is rather technical, requiring an explicit application of the axiom of choice. Our covariantisation map does not generalise Werner's, nor is it generalised by his, since ours is not formulated for observables with infinitely many outcomes, but covers examples where the symmetries are not those of a phase-space.

\Cref{sec:background-defns} contains background definitions. In \cref{sec:invariant-mean} we define the covariantisation map and use it to derive our central results. Sections \cref{sec:qubit-triple,sec:fourier-pair} are case studies, applying the previous results to the three mutually unbiased qubit observables, and phase space observables for an arbitrary finite state-space, respectively.

\section{Definitions and error measures}
\label{sec:background-defns}
\begin{figure}[ht]
  \centering
  \begin{tikzpicture}[scale=1]
    \tikzset{IdealMeasStyle/.style = {shape          = rectangle, rounded corners,
        fill     = blue!30,
        inner sep      = 5pt,
        font=\Huge
      }}
    \tikzset{ApproxMeasStyle/.style = {shape          = rectangle,
        fill     = orange!30,
        inner sep      = 5pt,
        font=\Huge
      }}
    \tikzset{JointMeasStyle/.style = {shape          = circle,
        fill     = green!30,
        inner sep      = 5pt,
        font=\Huge
      }}
    \tikzset{MarginEdgeStyle/.style   = {->, ultra thick,
        blue,
      }}
    \tikzset{ApproxEdgeStyle/.style   = {->,  >=latex,ultra thick,
        red,
        decorate,
        decoration = {snake,pre length=3pt,post length=7pt}
      }}
    \tikzset{node distance = 2}
    \node[IdealMeasStyle](E1){$\ope_1$};
    \node[IdealMeasStyle,below=of E1](E2){$\ope_2$};
    \node[IdealMeasStyle,below=of E2](En){$\ope_n$};
    \node[ApproxMeasStyle,right=of E1](F1){$\opf_1$};
    \node[ApproxMeasStyle,below=of F1](F2){$\opf_2$};
    \node[ApproxMeasStyle,below=of F2](Fn){$\opf_n$};
    \node[JointMeasStyle,right=of F2](J){J};
    \draw[ApproxEdgeStyle](F1) to node{} (E1) ;
    \draw[ApproxEdgeStyle](F2) to node{} (E2) ;
    \draw[ApproxEdgeStyle](Fn) to node{} (En) ;
    \draw[MarginEdgeStyle](J) to node{} (F1) ;
    \draw[MarginEdgeStyle](J) to node{} (F2) ;
    \draw[MarginEdgeStyle](J) to node{} (Fn) ;
    \path (E2) -- node[auto=false]{{\huge$\vdots$}} (En);
    \path (F2) -- node[auto=false]{{\huge$\vdots$}} (Fn);
  \end{tikzpicture}
  \caption{Target observables $\ope_i$,  compatible approximations $\opf_i$, and their joint $\opj$}
  \label{fig:depict-meas-uncertainty}
\end{figure}
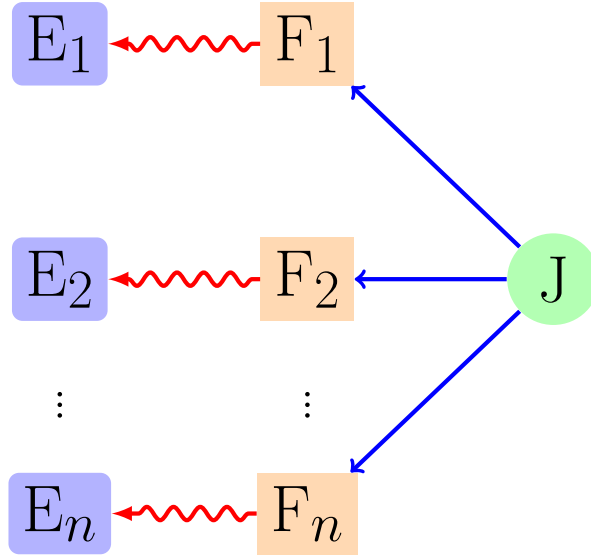

We consider separable complex Hilbert spaces, which we do not assume to be finite dimensional, and finite outcome observables. A map $\ope:\Omega\to \posops[\hcal]$ is an \emph{observable} if it is normalised
\begin{align}
  \sum_{\omega\in\Omega} \ope(\omega) &= \opi,
\end{align}
where $\opi$ is the identity operator on $\hcal$ and $\posops[\hcal]$ is the cone of positive operators.

A set of $n$ observables on the Hilbert space $\hcal$, $\sbuild{\ope_i\,}{\,i=1\hdots n}$, with outcome sets $\Omega_i$ is \emph{compatible} if there exists a joint observable, $\opj: \Omega \to \posops[\hcal]$,
\begin{align}
  \label{eqn:cart-joint}
  \sum_{\substack{\bm{\omega}\in\Omega\\\bm{\omega}_i = \omega^*}}\opj(\bm{\omega}) = \ope_i(\omega^*), \quad \forall i\in 1\hdots n,\ \forall \omega^*\in\Omega_i,
\end{align}
where $\Omega = \prod_i \Omega_i$ is the Cartesian product and $\bm{\omega}_i$ is the $i^\text{th}$ component of $\bm{\omega}$. Such an observable is called a Cartesian joint for the observables $\sbuild{\ope_i}{i\in 1\hdots n}$.

A linear map $R:\posops[\hcal]\to\posops[\hcal]$ is called a \emph{symmetry operation} if it is unital, completely positive and normal. An application of Wigner's theorem~\cite{wigner1931,wigner1960} demonstrates that all such maps are of the form
\begin{align}
  R\left[A\right] &= U^* A U,
\end{align}
where $U$ is either unitary or anti-unitary see, for example,~\cite{statistical-structure-quantum-holevo} for a complete exposition.

If $G$ is a group, $R_g$ a representation of $G$ by symmetry operations on $\hcal$, and $f_g:X\to X$ an action of $G$, we call $(G, R, f)$ a \emph{\covtrip}. An observable $A:X\to\posops[\hcal]$ is covariant with respect to $(G,R,f)$ if the relations
\begin{align}
  R_g\left[A(x)\right] = A(f_g(x)),
\end{align}
hold for all $g\in G$ and $x\in X$, in this case $(A, G, R, f)$ is known as a \emph{system of covariance}. This may be more or less restrictive depending on the choice of $(G,R,f)$, for example if $G$ is an arbitrary group, $R_g$ is the trivial representation $R_g\left[A(x)\right] = A(x)$, for all $g\in G$ and $f_g$ is the trivial action $f_g(x) = x$, then all observables are covariant. Less trivial examples are given in \cref{sec:qubit-triple,sec:fourier-pair}. Our notation for representations and actions is such that
\begin{align}
  f_e(x) &= x & f_{gh} &= f_g\circ f_h \\
  R_e[A] &= A & R_{gh} &= R_g\circ R_h,
\end{align}
in other words the map that takes $g$ to its corresponding element in the action or the representation is a group homomorphism. 
Given a pair of probability distributions, $\mu$ and $\nu$  over the same (finite) set $\Omega$, we can compute the $p$-norm of their (pointwise) difference. The resulting quantity is a metric on the space of probability distributions over $\Omega$,
\begin{align}
  \delta_p(\mu,\nu) &\defeq \pnorm{\mu - \nu}\\
                    &= \left(\sum_{\omega\in\Omega}\abs{\mu(\omega) - \nu(\omega)}^p\right)^{\frac{1}{p}}, \quad \forall p\in[1,\infty),\\
  \delta_\infty(S,T) &\defeq \max_{\omega\in\Omega}\abs{\mu(\omega) - \nu(\omega)}.
\end{align}
We note that $\delta_p(\mu,\nu) \geq 0$ with equality if and only if $\mu = \nu$ and that
\begin{align}
  \delta_p(\mu,\nu) \leq 2^{\frac{1}{p}}, \quad \forall p\in [1,\infty].
\end{align}
When $p=1$ this quantity is proportional to the total variation distance, and is also equal to the Wasserstein $1$-distance between $\mu$ and $\nu$, where the ``cost-function'' is given by the discrete metric \cite{markov-mixing-levin-peres-wilmer}{optimal-transport-villani}.
Given an observable $\ope:\Omega\to\posops[\hcal]$ and a quantum state $\rho\in\dops[\hcal]$ we can define a probability distribution over $\Omega$ via the Born rule,
\begin{equation}
  \ope^\rho:\omega \mapsto \tr{\ope(\omega)\rho}.
\end{equation}
We can lift the distance measure on probability distributions with outcome set $\Omega$ to one on observables on the same set, simply by taking the $\sup$ of the distance for the probability distributions over all states
\begin{align}
  d_p(\ope, \opf) &\defeq \sup_{\rho\in\dops[\hcal]} \delta_p(\ope^\rho, \opf^\rho)\\
                  &= \sup_{\rho\in\dops[\hcal]} \left(\sum_{\omega\in\Omega}\abs{\ope^\rho(\omega) - \opf^\rho(\omega)}^p\right)^{\frac{1}{p}},
\end{align}
where $\dops[\hcal]$ is the set of trace $1$ density operators within $\posops[\hcal]$. The supremum exists, because the expression is bounded. We define the measurement uncertainty region for this error measure to be the set
\begin{equation}
  S_p(\ope_1,\hdots \ope_n) = \sbuild{(d_p(\ope_1, \opf_1),\hdots d_p(\ope_n, \opf_n))}{\opf_i:\Omega_i\to\posops[\hcal] \text{ are compatible}},\quad p\in [1,\infty],
\end{equation} 
where $\ope_i:\Omega_i\to\posops[\hcal]$ are the target observables. It is this definition of the uncertainty region we will employ in the examples in \cref{sec:qubit-triple,sec:fourier-pair}, however the results of \cref{sec:invariant-mean} hold much more broadly. Specifically our results hold for error measures for observables obtained from real valued functions of pairs of probability distributions by taking the sup over the Born rule probability distributions. There are two conditions we require for the underlying error measure for probability distributions, firstly it must be jointly convex, in the sense that
\begin{align}\label{eqn:joint-convexity-condition}
  \delta(\lambda\nu_1 + (1-\lambda)\nu_2, \lambda\mu_1 + (1-\lambda)\mu_2) &\leq \lambda \delta_p(\nu_1, \mu_1) + (1-\lambda)\delta_p(\nu_2, \mu_2),
\end{align}
holds for $\lambda\in [0,1]$, and probability measures $\mu$,$\nu$. It must also be compatible with the \covtrip, specifically we require that
\begin{align}\label{eqn:compatibility-condition}
  \delta(\mu \circ f_g, \nu\circ f_g) = d(\mu,\nu),
\end{align}
where $\sbuild{f_g}{g\in G}$ are the elements of the group action. We argue that this condition is a natural one, if the action of the symmetry can change the error then it is reasonable to ask if the error measure is appropriate for studying the system in question. Although the condition is not satisfied for the Wasserstein distance under arbitrary symmetry actions, it is satisfied in the phase space examples studied in~\cite{6773660Werner:2004:URJ:2011593.2011606,PhysRevLett.111.160405,blw-meas-uncertainty,sharp-ur-num-angle,Werner2016}.

Other examples of error measures which satisfy our assumptions include a class of $f$-divergences which includes several ``divergences'' well known from the literature~\cite{1705001g}. All $f$-divergences have the joint-convexity property. Further, in the case where the probability measures are absolutely continuous with respect to a reference measure invariant under the group action the $f$-divergence will be compatible with the action. This is the case, for example, for probability distributions on finite sets, where the reference measure assigns the same probability to each element, and is therefore invariant under the action of the full symmetric group on the set.

\section{The covariantisation map}
\label{sec:invariant-mean}

It is convenient to embed the set of observables with a fixed outcome set $\Omega$ in the real vector space of all maps from $\Omega$ to the bounded, self adjoint operators on $\hcal$  which, for convenience, we denote $\ovms$. The set of such maps ranging in the positive operators is a convex cone in $\ovms$, and the normalisation $\sum_\omega\ope(\omega) = \opi$ defines an affine space. We equip the space $\ovms$ with a norm via
\begin{align}
  \norm{\ope} = \sum_{\omega\in\Omega} \norm{\ope(\omega)},
\end{align}
where $\norm{\ope(\omega)}$ denotes the operator norm.
\begin{defn}\label{defn:inv-mean}
  Given a covariance triple $\tau = (G, R, f)$ we define the \emph{covariantisation map}, $\cov[\tau]:\ovms\to\ovms$ by
  \begin{equation}\label{eqn:inv-mean-defn}
    \cov[\tau]{}[\ope](\omega) = \frac{1}{\abs{G}}\sum_{g\in G} R_{g^{-1}}\! \left[\ope(f_g(\omega))\right],
  \end{equation}
\end{defn}
We summarise some useful properties, in \cref{lem:inv-mean-contraction,lem:inv-mean-obs-to-obs,lem:inv-mean-proj}. 
\begin{lem}\label{lem:inv-mean-contraction}
  The invariant mean is a norm contraction.
  \begin{proof}
    \begin{align}
      \norm{\cov[\tau]{}[\ope]} &= \sum_{\omega\in\Omega} \norm{\cov[R][f][\ope](\omega)}\\
                           &= \sum_{\omega\in\Omega} \norm{\frac{1}{\abs{G}}\sum_{g\in G} R_{g^{-1}}\! \left[\ope(f_g(\omega))\right]}\\
                           &\leq \sum_{\omega\in\Omega} \frac{1}{\abs{G}}\sum_{g\in G} \norm{ R_{g^{-1}}\! \left[\ope(f_g(\omega))\right]}\\
                           &=\frac{1}{\abs{G}}\sum_{g\in G} \sum_{\omega\in\Omega}  \norm{ R_{g^{-1}}\! \left[\ope(f_g(\omega))\right]}\\
                           &=\frac{1}{\abs{G}}\sum_{g\in G} \sum_{\omega\in\Omega}  \norm{ \ope(f_g(\omega))} \label{eqn:inv-mean-norm-contraction-remove-rep} \\ 
                           &=\frac{1}{\abs{G}}\sum_{g\in G} \sum_{\omega\in\Omega}  \norm{ \ope(f_g(\omega))}\\
                           &\leq \norm{\ope},
    \end{align}
    where \cref{eqn:inv-mean-norm-contraction-remove-rep} follows from Wigner's theorem, and noting that (anti-)unitaries are norm preserving.
  \end{proof}
\end{lem}

\begin{lem}\label{lem:inv-mean-obs-to-obs}
  The invariant mean of an observable is an observable.
  \begin{proof}
    For any observable $\ope:\Omega\to\posops[\hcal]$ the map $\cov[\tau]{}[\ope]$ takes positive values since the $R_g$ are positive, and the positive operators form a convex set. Further, if $\ope$ is an observable then so is $\cov[\tau]{}[\ope]$, since
    \begin{align}
      \sum_{\omega\in\Omega} \cov[R][f][\ope](\omega) &= \frac{1}{\abs{G}} \sum_{\omega\in\Omega} \sum_{g\in G} R_{g^{-1}}\! \left[\ope(f_g(\omega))\right]\\
                                                   &= \frac{1}{\abs{G}} \sum_{g\in G} R_{g^{-1}}\! \left[\sum_{\omega\in\Omega}\ope(f_g(\omega))\right]\\
                                                   &= \frac{1}{\abs{G}} \sum_{g\in G} \opi\\
                                                   &= \opi.
    \end{align}
  \end{proof}
\end{lem}
\begin{lem}\label{lem:inv-mean-proj}
  The invariant mean is the projection from $\ovms$ onto the subspace of $(R,f)$-covariant maps.
  \begin{proof}
    First note that $\cov[\tau]{}$ is linear, since the $R_g$ are linear. For any $\ope\in\ovms$, $\cov[\tau]{}[\ope]$ is covariant since
    \begin{align}
      \cov[\tau]{}[\ope](f_h(\omega)) &= \frac{1}{\abs{G}} \sum_{g\in G} R_{g^{-1}}\! \left[\ope(f_g\circ f_h(\omega))\right]\\
                                 &= \frac{1}{\abs{G}} \sum_{g^\prime\in G} R_{{(g^\prime h^{-1})}^{-1}}\! \left[\ope(f_{g^\prime h^{-1}}\circ f_h(\omega))\right]\\
                                 &= \frac{1}{\abs{G}} \sum_{g^\prime\in G} R_{h{g^\prime}^{-1}} \! \left[\ope(f_{g^\prime}(\omega))\right]\\
                                 &= R_h\!\left[\frac{1}{\abs{G}} \sum_{g^\prime\in G} R_{{g^\prime}^{-1}} \! \left[\ope(f_{g^\prime}(\omega))\right]\right]\\
                                 &= R_h\!\left[\cov[\tau]{}[\ope](\omega)\right].
    \end{align}
    Now $(R,f)$-covariant maps are invariant under $\cov[\tau]{}$
    \begin{align}
      \cov[\tau]{}[\ope](\omega) &= \frac{1}{\abs{G}}\sum_{g\in G} R_{g^{-1}}\! \left[\ope(f_g(\omega))\right]\\
                            &= \frac{1}{\abs{G}}\sum_{g\in G} R_{g^{-1}}\! \left[R_{g}\!\left[\ope(\omega)\right]\right]\\
                            &= \frac{1}{\abs{G}}\sum_{g\in G} \ope(\omega)\\
                            &= \ope(\omega),
    \end{align}
    so $\cov[\tau]{}$ is idempotent.
  \end{proof}
\end{lem}
It follows that the space of $(G,R,f)$-covariant maps is a vector subspace of $\ovms$, and that $\ope=\cov[\tau]{}[\ope]$ if, and only if, $\ope$ is $(G, R,f)$-covariant.

\begin{thm}
  \label{thm:inv-mean-reduces-error}
  Let $\tau = (G, R_g, f_g)$ be a \covtrip{} and let $d$ be an error measure for probability distributions, which is jointly convex, and compatible with $\tau$ in the sense defined in \cref{eqn:compatibility-condition}, then
  \begin{align}
    \sup_\rho d\left(\cov[\tau]{}[\ope]^\rho, \cov[\tau]{}[\opf]^\rho\right) \leq \sup_\rho d\left(\ope^\rho,\opf^\rho\right),
  \end{align}
  For all observables $\ope, \opf:\Omega\to\posops[\hcal]$.
  \begin{proof}
    \begin{align}
      \sup_\rho d\left(\cov[\tau]{}[\ope]^\rho, \cov[\tau]{}[\opf]^\rho\right) &= \sup_\rho d\left( \frac{1}{\abs{G}}\sum_{g\in G} \left(R_{g^{-1}} \circ \ope\circ f_g\right)^\rho , \frac{1}{\abs{G}}\sum_{g\in G}\left(R_{g^{-1}} \circ \opf\circ f_g\right)^\rho \right)\\
                                                                     &\leq \sup_\rho \frac{1}{\abs{G}}\sum_{g\in G} d\left( \left(R_{g^{-1}} \circ \ope\circ f_g\right)^\rho ,\left(R_{g^{-1}} \circ \opf\circ f_g\right)^\rho \right)\\
                                                                     &\leq \frac{1}{\abs{G}}\sum_{g\in G}\sup_\rho d\left( \left(R_{g^{-1}} \circ \ope\circ f_g\right)^\rho ,\left(R_{g^{-1}} \circ \opf\circ f_g\right)^\rho \right)\\
                                                                     &= \frac{1}{\abs{G}}\sum_{g\in G}\sup_\rho d\left( \left( \ope\circ f_g\right)^{R_{g^{-1}}^*[\rho]} ,\left(\opf\circ f_g\right)^{R_{g^{-1}}^*[\rho]} \right)\\
                                                                     &\leq \frac{1}{\abs{G}}\sum_{g\in G}\sup_\rho d\left( \left( \ope\circ f_g\right)^{\rho} ,\left(\opf\circ f_g\right)^{\rho} \right)\\
                                                                     &= \frac{1}{\abs{G}}\sum_{g\in G}\sup_\rho d\left( \ope^{\rho}\circ f_g , \opf^{\rho}\circ f_g \right)\\
                                                                     &= \frac{1}{\abs{G}}\sum_{g\in G}\sup_\rho d\left(\ope^{\rho} ,\opf^{\rho}\right)\\
                                                                     &= \sup_\rho d\left(\ope^{\rho} ,\opf^{\rho}\right).
    \end{align}
  \end{proof}
\end{thm}

Given a set of $n\in\mathbb{N}$ finite sets $\sbuild{\Omega_i}{i\in 1\hdots n}$, and $n$ finite groups $\sbuild{G_i}{i\in 1\hdots n}$, with action $f_{g_i}^i: \Omega_i \to \Omega_i$, for each $g_i\in G_i$ there is a \emph{product action} $\pi$ of the direct product group $G = \prod_i G_i$ on the Cartesian product set $\Omega = \prod_i \Omega_i$
\begin{align}
  \pi_g&: \Omega \to \Omega, \quad \forall g\in G\\
  \pi_{(g_1, \hdots, g_i,\hdots, g_n)} &: (\omega_1, \hdots \omega_n) \mapsto (f_{g_1}^1(\omega_1), \hdots, f_{g_n}^n(\omega_n)),
\end{align}
there is also a marginal action $\mu^i$ of the direct product group on each $\Omega_i$
\begin{align}
  \mu^i_g &:\Omega_i \to \Omega_i, \quad \forall g\in G\\
  \mu^i_{(g_1,\hdots, g_i,\hdots,g_n)}&:\omega\mapsto f^i_{g_i}(\omega).
\end{align}

\begin{lem}\label{lem:margins-inv-mean-same}
  For $i\in 1\hdots n$ let $\ope_i:\Omega_i \to \posops[\hcal]$ be a compatible family of observables, and $\sbuild{G_i}{i\in 1\hdots n}$ be a set of groups, such that $G_i$ has action $f_g^i$ on $\Omega_i$. Let $\Omega = \prod_i \Omega_i$ be the Cartesian product, $G = \prod_i G_i$ the direct product, and $\pi$, $\mu^i$ the product and marginal actions of $G$ respectively. Let $\sbuild{R_g}{g\in G}$ be a representation of $G$ as positive, unital, linear maps acting on $\saops[\hcal]$. If $\opj$ is any Cartesian joint observable for the $\ope_i$, and $\tilde{\opj}_i$ the $i^\text{th}$ margin of $\cov[R][\pi][\opj]$,
  \begin{align}
    \tilde{\opj}_i&: \Omega_i \to \posops[\hcal]\\
    \tilde{\opj}_i &: \omega^* \mapsto \sum_{\substack{\underline{\omega} \in \Omega\\ \underline{\omega}_i = \omega^*}}\cov[R][\pi][\opj](\underline{\omega}),
  \end{align}
  where $\underline{\omega}_i$ denotes the $i^\text{th}$ element of the tuple $\underline{\omega}$, then
  \begin{align}
    \tilde{\opj}_i(\omega) &= \cov[R][\mu^i][\ope_i](\omega), \quad \forall \omega \in \Omega_i.
  \end{align} 
  \begin{proof}
    \begin{align}
      \tilde{\opj}_i(\omega^*) &= \sum_{\substack{\underline{\omega} \in \Omega\\ \underline{\omega}_i = \omega^*}}\cov[R][\pi][\opj](\underline{\omega})\\
                               &= \sum_{\substack{\underline{\omega} \in \Omega\\ \underline{\omega}_i = \omega^*}} \frac{1}{\abs{G}}\sum_{g\in G} R_{g^{-1}}\left[\opj(\pi_g(\underline{\omega}))\right]\\
                               &= \frac{1}{\abs{G}}\sum_{g\in G} R_{g^{-1}}\left[\sum_{\substack{\underline{\omega} \in \Omega\\ \underline{\omega}_i = \omega^*}} \opj(\pi_g(\underline{\omega}))\right]\\
                               &= \frac{1}{\abs{G}}\sum_{g\in G} R_{g^{-1}}\left[\sum_{\substack{\underline{\omega} \in \Omega\\ \pi_{g^{-1}}(\underline{\omega})_i = \omega^*}} \opj(\underline{\omega})\right]\\
                               &= \frac{1}{\abs{G}}\sum_{g\in G} R_{g^{-1}}\left[\sum_{\substack{\underline{\omega} \in \Omega\\ \underline{\omega}_i = \mu^i_{g}(\omega^*)}} \opj(\underline{\omega})\right]\\
                               &= \frac{1}{\abs{G}}\sum_{g\in G} R_{g^{-1}}\left[\ope\left(\mu^i_{g}(\omega^*)\right)\right]\\
                               &= \cov[R][\mu^i][\ope_i](\omega^*).
    \end{align}
  \end{proof}
\end{lem}

\begin{figure}[ht]
  \centering
  \begin{tikzpicture}
    \matrix (m) [matrix of math nodes,row sep=6em,column sep=6em, font=\Huge]
    {
      J & \tilde{J} \\
      E_i & \tilde{E}_i \\};
    \path[->] (m-1-1) edge node [font=\huge,left] {$M_i$} (m-2-1);
    \path[->] (m-1-1) edge node [font=\huge,above] {$\cov[R][\pi]{}$} (m-1-2);
    \path[->] (m-2-1) edge node [font=\huge,below] {$\cov[R][\pi_i]{}$} (m-2-2);
    \path[->] (m-1-2) edge node [font=\huge,right] {$M_i$} (m-2-2);
  \end{tikzpicture}
  \caption{An illustration of \cref{lem:margins-inv-mean-same}: starting from the approximating joint observable one can either apply the full invariant mean $\cov[R][\pi]$ and then take the margins with the $M_i$, or first marginalise with the $M_i$ and then apply the invariant means $\cov[R][\mu_i]$.}
  \label{fig:picture-of-lemma-4}
\end{figure}

\begin{thm}
  \label{thm:cov-obs-improve-approx}
  Let $\sbuild{\ope_i}{ i\in 1\hdots n}$ be a family of (not necessarily compatible) observables, $\ope_i:\Omega_i \to \posops[\hcal]$, and $\sbuild{G_i}{i\in 1\hdots n}$ be a set of groups, such that $G_i$ has action $f_g^i$ on $\Omega_i$. Let $\Omega = \prod_i \Omega_i$ be the Cartesian product, $G = \prod_i G_i$ the direct product, and $\pi$, $\mu^i$ the product and marginal actions of $G$ respectively. Let $\sbuild{R_g}{g\in G}$ be a representation of $G$ as positive, unital, linear maps acting on $\saops[\hcal]$ such that
  \begin{align}
    \cov[R][\mu^i][\ope_i] = \ope_i.
  \end{align}
  Then for any compatible family of observables $\sbuild{\opf_i}{ i\in 1\hdots n}$, $\opf_i:\Omega_i \to \posops[\hcal]$, with joint observable $\opj:\Omega\to\posops[\hcal]$, the observables
  \begin{align}
    \tilde{\opf}_i &= \cov[R][\mu^i][\opf_i]
  \end{align}
  are compatible, with joint $\tilde{\opj} = \cov[R][\pi][\opj]$, and for any function $d$ satisfying the constraints of \cref{thm:inv-mean-reduces-error}
  \begin{align}
    \label{eqn:inv-mean-reduce-error}
    \sup_\rho d(\tilde{\opf}^\rho_i, \ope_i^\rho) \leq \sup_\rho d(\opf_i^\rho, \ope_i^\rho).
  \end{align}
  \begin{proof}
    The compatibility of the $\tilde{\opf}_i$ follows directly from \cref{lem:margins-inv-mean-same}, therefore it only remains to establish inequality \eqref{eqn:inv-mean-reduce-error},
    \begin{align}
      \sup_\rho  d(\tilde{\opf}_i, \ope_i) &= \sup_\rho d(\cov[R][\mu^i][\opf_i], \ope_i)\\
      \label{eqn:line-target-inv}
                                           &= \sup_\rho d(\cov[R][\mu^i][\opf_i],  \cov[R][\mu^i][\ope_i])\\
      \label{eqn:line-inv-mean-contraction}
                                           &\leq d(\opf_i, \ope_i).
    \end{align}
    \Cref{eqn:line-target-inv} is a consequence of assuming the target observables are unchanged by the invariant mean, and \eqref{eqn:line-inv-mean-contraction} is the result of \cref{thm:inv-mean-reduces-error}.
  \end{proof}
\end{thm}

It is tempting to attempt to generalise \cref{eqn:inv-mean-defn}. For a locally compact group $G$, with (left) Haar measure $\mu$, continuous action $\alpha: (g,\omega)\mapsto g\cdot\omega$ on a Borel measurable, locally compact space $(\Omega, \fcal)$, and continuous representation $R_g$, on $\posops[\hcal]$ one might try to define
\begin{align}
  M[F]: X\mapsto\int_G d\mu(g) R_g[F(g.X)],
\end{align}
for a POVM $F:\fcal\to\posops[\hcal]$. Unfortunately there are significant technical obstacles to defining such a quantity. In particular one would have to show the function $g\mapsto R_g[F(g.X)]$ is $\mu$-measurable, in the sense of the Bochner integral \cite{measure-theory-cohn}, either for all observables, or for a physically relevant subset. In the (possibly highly restricted) cases that such a quantity may be defined it is easy to see that it will be necessary for the group $G$ to be compact, rather than locally compact since
\begin{align}
  M[F]:\Omega&\mapsto\int_G d\mu(g) R_g[F(g.\Omega)]\\
             &=\int_G d\mu(g) R_g[\opi]\\
             &=\opi\int_G d\mu(g).\label{eqn:inv-mean-identity-infinite-case}
\end{align}
The Haar measure $\mu$ may be normalised to a probability measure if, and only if, the group is compact. This excludes several physically relevant groups including the translation group of $\R$ or $\R^n$, the Galilei group and the Poincar{\'e} group. Compact groups relevant to physical applications include the finite groups covered above, the unitary, special unitary groups, orthogonal and special orthogonal groups in $n\in\N$ dimensions. 

The generalised invariant mean will require additional regularity conditions on the observables it is applied to. To see why this is the case we recall that to be Bochner integrable the function $g\mapsto R_g[F(g.X)]$ must be the limit of piecewise constant functions, where the pieces are measurable sets. To take a concrete example we restrict our attention to probability measures on the circle. Let, $\Omega=[-\pi,\pi)$, $(\Omega, \tau)$ be the topological space of the unit circle, and let $X=(0,1)\subset\Omega$, $F:\borel{\Omega}[\tau]\to [0,1]$ be the point measure, assigning $1$ to sets if they contain the element $0$, and $0$ otherwise otherwise. Finally take $G$ to be the circle group and $f_g:h\mapsto gh$ be the action of the circle group on itself. Consider a sequence $g_n$ of negative elements of $\Omega$, converging to zero, then
\begin{align}
  F(f_{g_n}(X)) &= F(g_n + X)\\
                &= F( (g_n, 1+g_n))\\
                &= 1,
\end{align}
whereas $F(X) = 0$. With general observables it is difficult to control these discontinuities. We conjecture that a necessary and sufficient condition for measurability is the existence of a covariant observable dominating $F$.

For completeness we draw attention to the key limitation of \cref{lem:margins-inv-mean-same,thm:cov-obs-improve-approx}, for simplicity in the case of two target observables. It is not the case that \cref{thm:cov-obs-improve-approx} applies to an arbitrary pair of systems of covariance $(G_1, R_g^1, f_g^1, \ope_1)$ and $(G_2, R_g^2, f_g^2, \ope_2)$. Instead we take the direct product $G = G_1\times G_2$ and use a representation $R$ of the product group in both systems. In practice this means that if one wishes to apply this theorem to two systems of covariance it is necessary that they are compatible in some sense. A sufficient condition is that the effects of each observable are unchanged by all of the elements of the representation associated with the other, and that the elements of each representation commute with all of the elements of the other.

\section{Pauli observables}
\label{sec:qubit-triple}
Let $\boa$, $\bob$ and $\boc$ be three orthonormal vectors in $\mathbb{R}^3$, and consider the three, two outcome qubit observables
\begin{align}
  \opa&:\{+1, -1\} \to \mathcal{L}\left(\mathbb{C}^2\right), & \opb&:\{+1, -1\} \to \mathcal{L}\left(\mathbb{C}^2\right), & \opc&:\{+1, -1\} \to \mathcal{L}\left(\mathbb{C}^2\right)\\
  \opa&:k\mapsto \qbit{k\boa}, &\opb&:l\mapsto \qbit{l\bob}, &\opc&:m\mapsto \qbit{m\boc}.
\end{align}
We would like to find the set
\begin{align}
  S(\opa, \opb, \opc) &= \sbuild{(d(\opa, \opd), d(\opb, \ope), d(\opc, \opf))}{\opd, \ope, \opf:\B^3\to\mathcal{L}\left(\mathbb{C}^2\right) \text{ are compatible}},
\end{align}
where $\B = \{-1,1\}$. The condition that $\opd$, $\ope$, $\opf$ are compatible is equivalent to the existence of an observable $\opj:\B^3\to\mathcal{L}\left(\mathbb{C}^2\right)$ such that
\begin{align}
  \sum_{l,m}\opj(k,l,m) &= \opd(k)\\
  \sum_{k,m}\opj(k,l,m) &= \ope(l)\\
  \sum_{k,l}\opj(k,l,m) &= \opf(m).
\end{align}

Since we have three, two outcome target observables we take as our product group the elementary Abelian group of order $8$, the additive group of the vector space $\left(\mathbb{Z}/2\mathbb{Z}\right)^3$
\begin{align}
  G &= \sbuild{\grp{k}{l}{m}}{(k,l,m)\in \{+1, -1\}^3}\\
  \grp{h}{i}{j} \grp{k}{l}{m} &= \grp{hk}{il}{jm}.
\end{align}
This group has product action on the outcome set $\{+1,-1\}^3$
\begin{equation}
  \pi_{h,i,j}\left((k,l,m)\right) = (hk,il,jm),
\end{equation}
and marginal actions
\begin{align}
  \mu^1_{h,i,j} \left(k\right) &= hk\\
  \mu^2_{h,i,j} \left(l\right) &= il\\
  \mu^3_{h,i,j} \left(m\right) &= jm.
\end{align}
It may be represented by the following set of positive, unital, linear maps on $\saops[\mathbb{C}^2]$
\begin{align}
  \map{k}{l}{m}\left[\qbit[r_0]{\begin{pmatrix}r_1\\r_2\\r_3\end{pmatrix}}\right] &= \qbit[r_0]{\begin{pmatrix}k r_1\\l r_2\\m r_3\end{pmatrix}}.
\end{align}

Given any compatible, two outcome qubit observables, $\opd$, $\ope$ and $\opf$, we can apply the invariant mean with respect to this group, action and representation to the joint $\opj$
\begin{equation}
  \tilde{\opj}(k,l,m) = \frac{1}{8}\sum_{(h,i,j)\in\{+1,-1\}^3} \map{h}{i}{j}\left[\opj(hk, il, jm) \right],
\end{equation}
and take the margins of $\tilde{\opj}$ to get a new set of compatible, two outcome qubit observables
\begin{align}
  \tilde{\opd}(k) &= \sum_{(l,m)\in\{+1,-1\}^2}\tilde{\opj}(k,l,m)\\
  \tilde{\ope}(l) &= \sum_{(k,m)\in\{+1,-1\}^2}\tilde{\opj}(k,l,m)\\
  \tilde{\opf}(m) &= \sum_{(k,l)\in\{+1,-1\}^2}\tilde{\opj}(k,l,m).
\end{align}
By \cref{lem:margins-inv-mean-same} this is equivalent to taking the invariant mean with respect to the $G$, $R$ and $\mu^i$ of $\opd$, $\ope$, $\opf$ directly
\begin{align}
  \tilde{\opd}(k) &= \cov[R][f^1][\opd](k) & \tilde{\ope}(l) &= \cov[R][f^2][\ope](l) & \tilde{\opf}(m) &= \cov[R][f^3][\opf](m)
\end{align}
These marginal groups, actions and representations satisfy all of the requirements of \cref{thm:cov-obs-improve-approx} above, so the group averaging maps reduce the error. We also have that each target observable is invariant under the respective $\cov[R][\mu^i]$. so we can apply \cref{thm:cov-obs-improve-approx} implying that for every compatible triple $\opd$, $\ope$, $\opf$ there exists a \emph{covariant} compatible triple with lower distances. Since we can also increase the distances by \cref{lem:infty-increase-error} as needed we can fill the set $S(\opa, \opb, \opc)$ by searching over the covariant observables, and then increasing the distances up to the trivial maximum of $2^{\frac{1}{p}}$.
The covariant joints have the form
\begin{align}
  \opj(k,l,m) = \qbit[][+][\frac{1}{8}]{\begin{pmatrix}k j_1\\ l j_2 \\ m j_3\end{pmatrix}},
\end{align}
for $\norm{\boj}\leq 1$. The margins have the form
\begin{align}
  \opd(k) &= \qbit{j_1 \boa} &
                               \ope(l) &= \qbit{j_2 \bob}&
                                                           \opf(m) &= \qbit{j_3 \boc},
\end{align}
which have distances
\begin{align}
  d_p(\opa, \opd) &= 2^{\frac{1}{p} -1}\left(1-j_1\right)&
                                                           d_p(\opb, \ope) &= 2^{\frac{1}{p} -1}\left(1-j_2\right)&
                                                                                                                    d_p(\opc, \opf) &= 2^{\frac{1}{p} -1}\left(1-j_3\right).
\end{align}
Hence the positivity constraint $\norm{\boj}\leq 1$ becomes
\begin{equation}
  \left(d(\opa, \opd) - 2^{\frac{1}{p}-1}\right)^2 + \left(d(\opb, \ope) - 2^{\frac{1}{p}-1}\right)^2 +\left(d(\opc, \opf) - 2^{\frac{1}{p}-1}\right)^2 \leq 2^{\frac{2}{p}-2}.
\end{equation}
The subset of $S_p$ covered by covariant observables is a sphere of radius $2^{\frac{1}{p}-1}$ centered at $\left(2^{\frac{1}{p}-1},2^{\frac{1}{p}-1},2^{\frac{1}{p}-1}\right)$, the full region is the monotone closure of this within the cube $[0, 2^{\frac{1}{p}}]^3$.

\begin{figure*}[t]
  \centering  
  \begin{subfigure}[t]{0.4\textwidth}
    \includegraphics[width=0.8\textwidth]{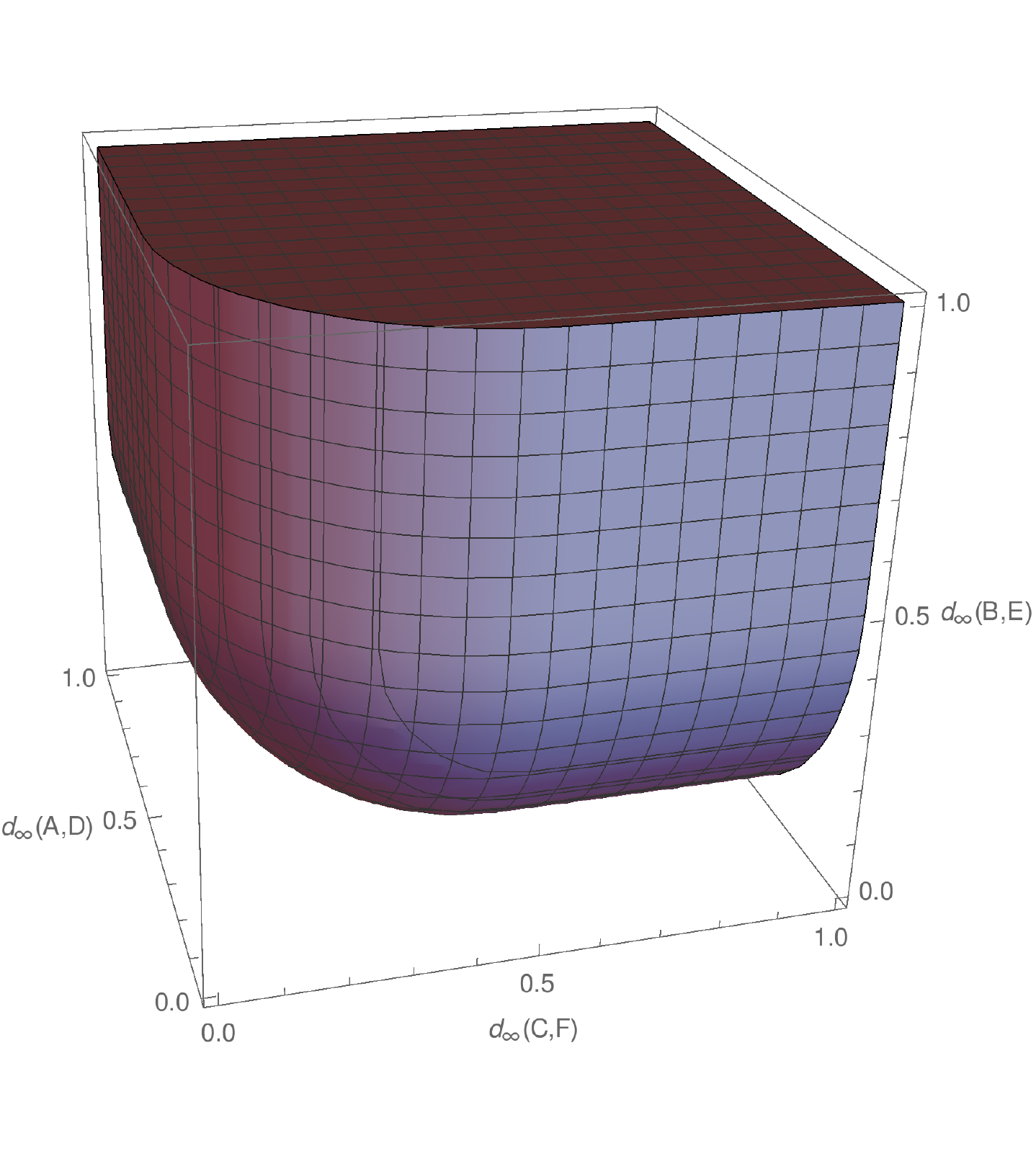}
  \end{subfigure}\quad
  \begin{subfigure}[t]{0.4\textwidth}
    \includegraphics[width=0.8\textwidth]{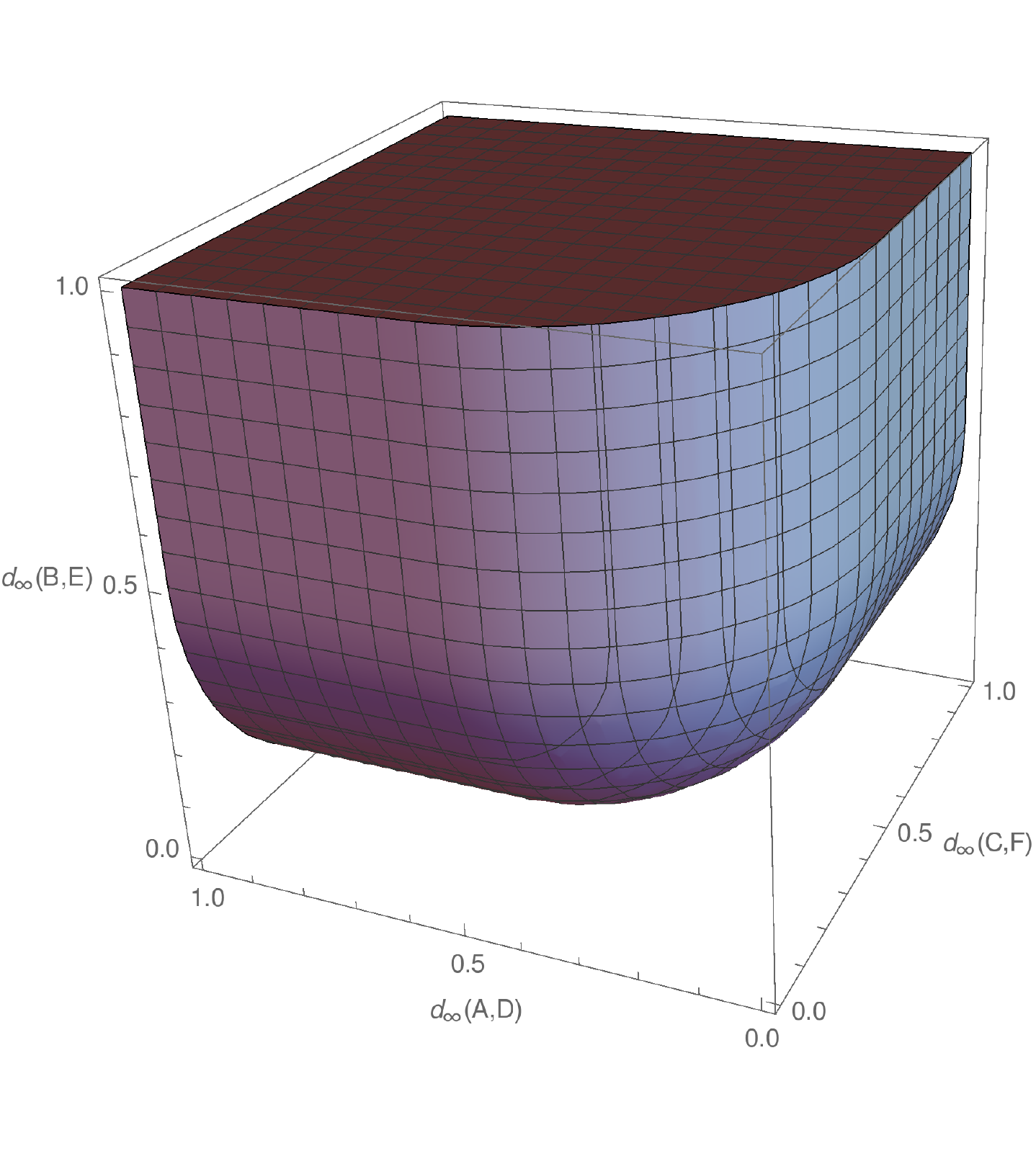}
  \end{subfigure}\\
  \begin{subfigure}[t]{0.4\textwidth}
    \includegraphics[width=0.8\textwidth]{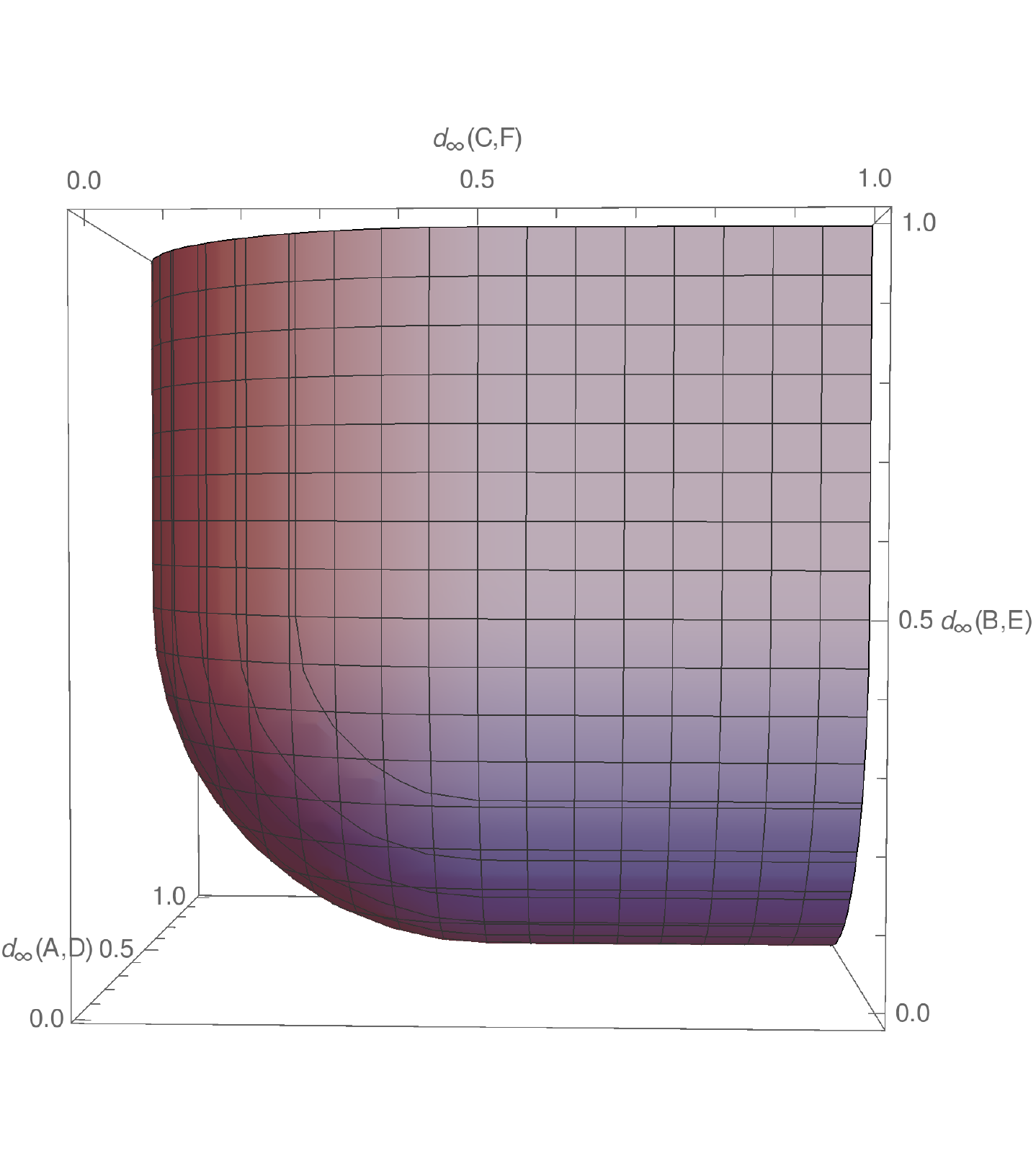}
  \end{subfigure}\quad
  \begin{subfigure}[t]{0.4\textwidth}
    \includegraphics[width=0.8\textwidth]{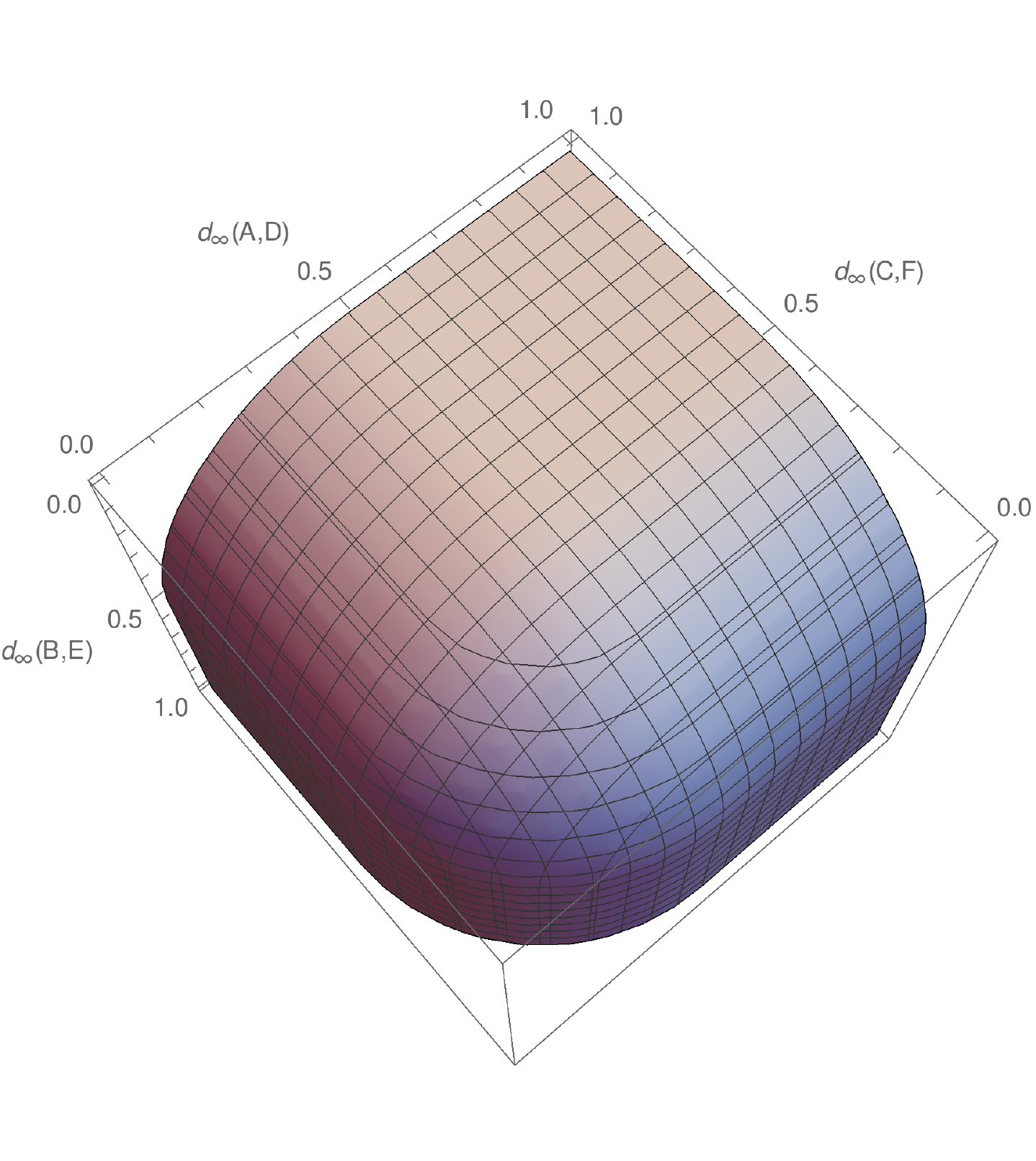}
  \end{subfigure}
  \caption{Views of the uncertainty region $S_\infty(\opa.\opb,\opc)$ covered by compatible approximations $\opd$, $\ope$ and $\opf$.}
\end{figure*}

\section{Finite phase space observables}
\label{sec:fourier-pair}
Let $\cyc[n] = \{0\hdots n-1\}$ denote the cyclic group of order $n$, equivalent to the set of natural numbers less than $n$, with the group operation addition modulo $n$, denoted $+$. Although this is only a field for $n$ prime, it will be useful to define multiplication, denoted by juxtaposition, as the usual multiplication of natural numbers modulo $n$. 

Let $\hcal$ be a Hilbert space of dimension $n\in\mathbb{N}$, $n \geq 2$, $\sbuild{\ket{g}}{g \in \cyc[n]}$ be an orthonormal set of vectors, hereafter called the \emph{computational basis} and let
\begin{align}
  \ket{f_h} &\defeq \sqrt{\frac{1}{n}}\sum_{g\in \cyc[n]} e^{\frac{2\pi i}{n} gh}\ket{g}, \quad h\in\cyc[n]\\
  \implies \ket{g} &= \sqrt{\frac{1}{n}}\sum_{h\in \cyc[n]} e^{-\frac{2\pi i}{n} gh} \ket{f_h}, \quad g\in\cyc[n].
\end{align}
The two bases are related by the well known quantum Fourier transform. It is easily verified that the $\ket{f_h}$ are an orthonormal basis for $\hcal$ and are \emph{mutually unbiased} with the computational basis. We define sharp observables for these bases
\begin{align}
  \opa&:\cyc[n] \to \posops[\hcal] & \opb&:\cyc[n] \to \posops[\hcal]\\
  \opa&:g\mapsto \ketbra{g}{g} & \opb&:h\mapsto \ketbra{f_h}{f_h}.
\end{align}
We can define unitary \emph{shift operators} for these bases
\begin{align}
  U_k \ket{g} &= \ket{g + k} & \forall g,k\in\cyc[n]\\
  V_q \ket{f_h} &= \ket{f_{h+q}}&\forall h,q\in\cyc[n],
\end{align}
and note that each form a unitary representation of the group $\cyc[n]$. Further, we have that
\begin{align}
  U_k &= \sum_{h\in\cyc[n]}e^{-\frac{2\pi i}{n} k h}\ketbra{f_h}{f_h} = \sum_{h\in\cyc[n]}e^{-\frac{2\pi i}{n} k h}\,\opb(h)\\
  V_q &= \sum_{g\in\cyc[n]}e^{\frac{2\pi i}{n} q g}\ketbra{g}{g} = \sum_{g\in\cyc[n]}e^{\frac{2\pi i}{n} q g}\,\opa(g).
\end{align}
One can verify the commutation relations
\begin{align}
  U_k V_q  &= e^{\frac{2\pi i}{n} kq}  V_q U_k,
\end{align}
by, for example, applying the operator on each side of the equality to the states in the Fourier basis. Therefore
\begin{align}
  U_k V_q \rho\, V_q^\dagger U^\dagger_k = V_q U_k \rho\,  U^\dagger_k V_q^\dagger, \quad \forall \rho\in\saops[\hcal].
\end{align}
We therefore consider the linear maps
\begin{align}
  R_{k,q}&:\saops[\hcal]\to \saops[\hcal]\\
  R_{k,q}&:\rho \mapsto U_k V_q \rho\, V^\dagger_q U^\dagger_k =  V_q U_k  \rho\,  U^\dagger_k V^\dagger_q,
\end{align}
and note that they form a representation of the direct product group $\cyc[n]\times\cyc[n]$, with the group operation given by operator composition
\begin{align}
  R_{k,q} \circ R_{l,r} = R_{k+l, p+r}, \quad \forall k,l,q,r\in \cyc[n].
\end{align}
These maps act on the effects of the target observables as
\begin{align}
  R_{k,q}\left[\ketbra{g}{g}\right] &=  \ketbra{g+k}{g+k}\\
  R_{k,q}\left[\ketbra{f_h}{f_h}\right] &=  \ketbra{f_{h+q}}{f_{h+q}}.
\end{align}

Therefore we can apply the methods of \cref{sec:invariant-mean} to establish that choosing covariant observables does not increase the error according to the $d_p$ distance measures. We refer to \cref{sec:app:computing-fourier-ur} for details of the calculation of the uncertainty region for covariant observables. The result is that the uncertainty region in dimension $n$ is the monotone closure of the ellipse which is tangent to each axis at coordinate $1-\frac{1}{n}$, and which has major axis along to the line $d_b = 1-\frac{1}{n}-d_a$.

\begin{figure*}
  \centering
  \begin{subfigure}[t]{0.4\textwidth}
    \includegraphics[width=\textwidth]{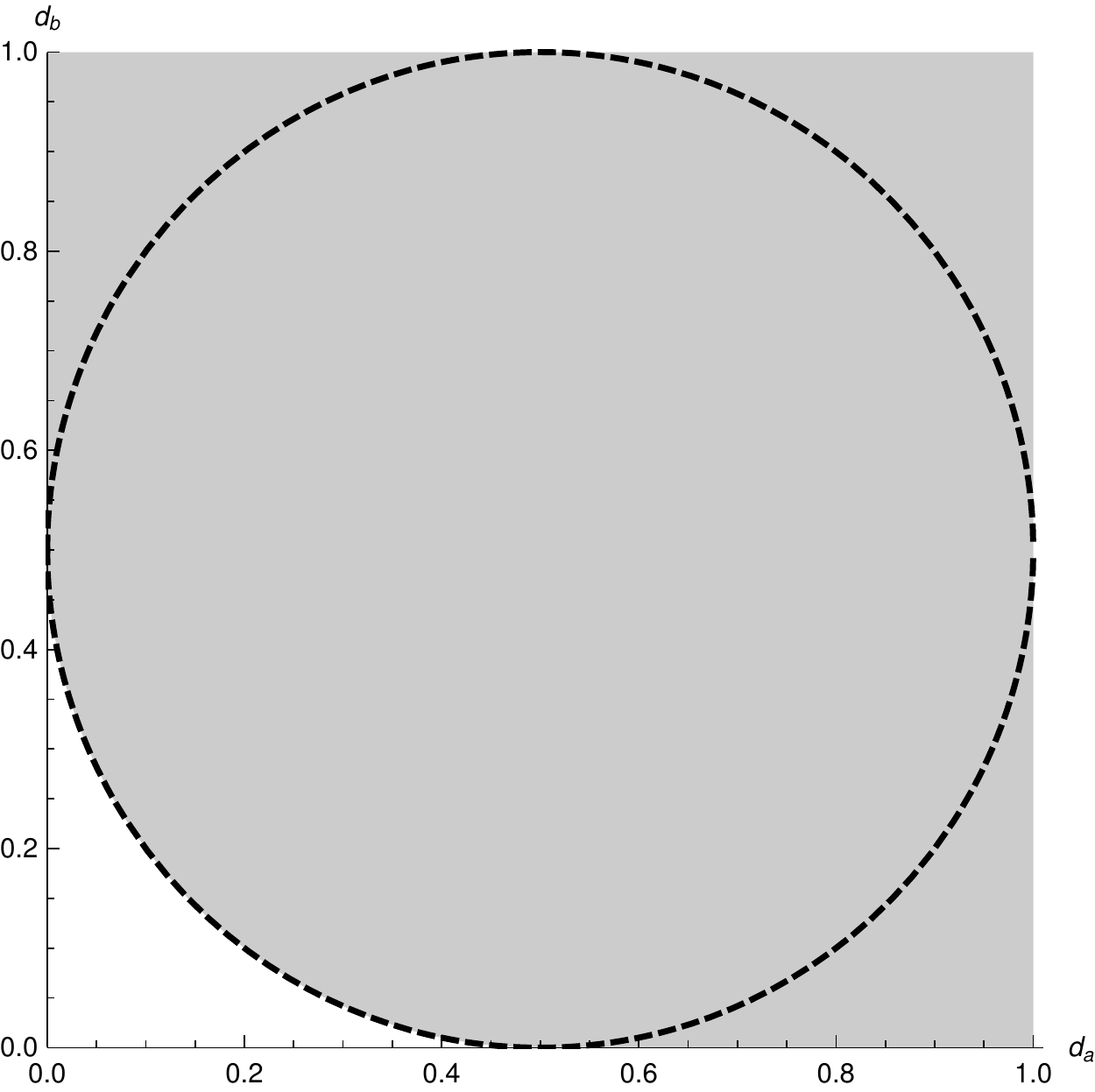}
  \end{subfigure}\quad
  \begin{subfigure}[t]{0.4\textwidth}
    \includegraphics[width=\textwidth]{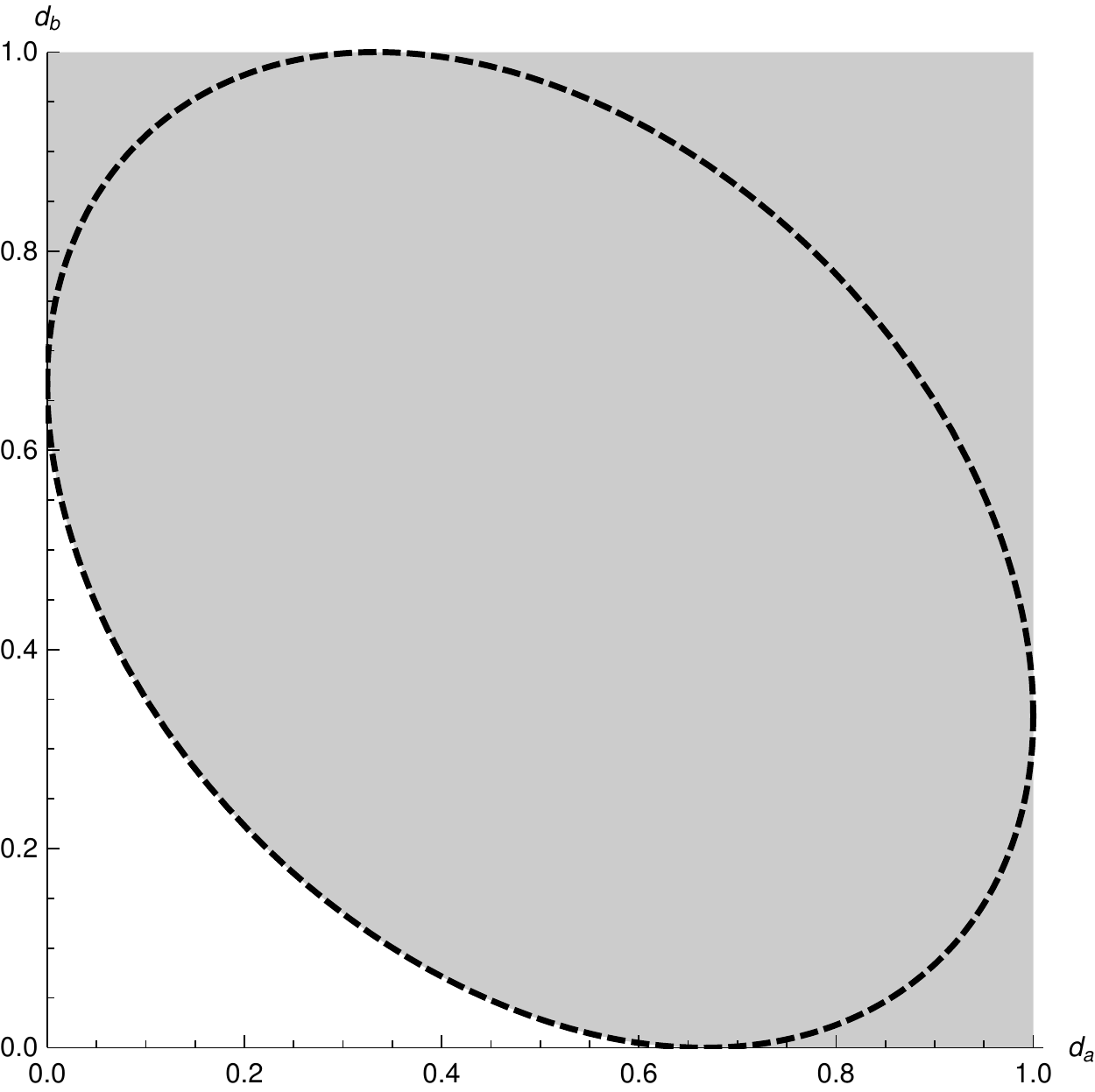}
  \end{subfigure}\\
  \begin{subfigure}[t]{0.4\textwidth}
    \includegraphics[width=\textwidth]{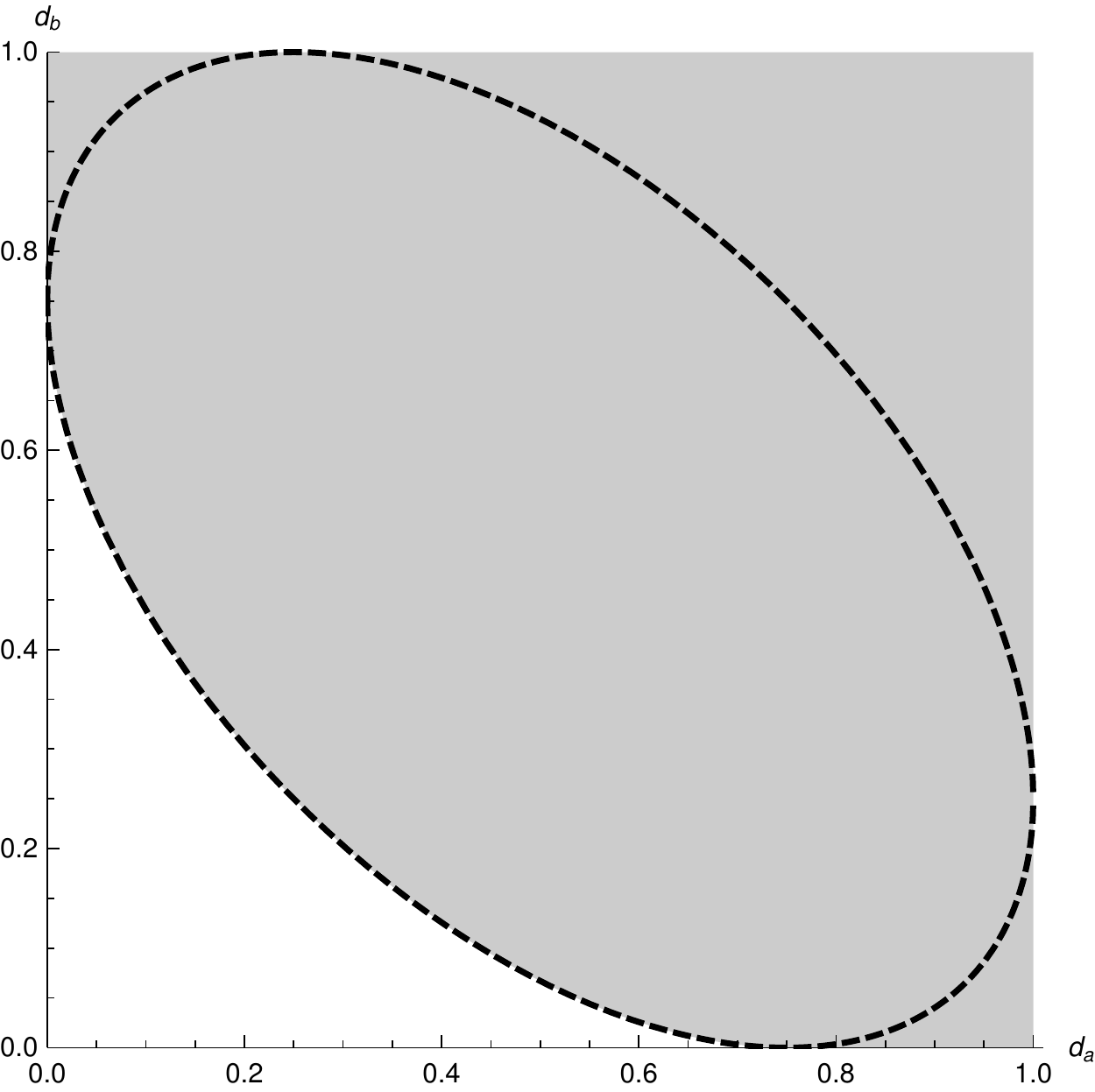}
  \end{subfigure}\quad
  \begin{subfigure}[t]{0.4\textwidth}
    \includegraphics[width=\textwidth]{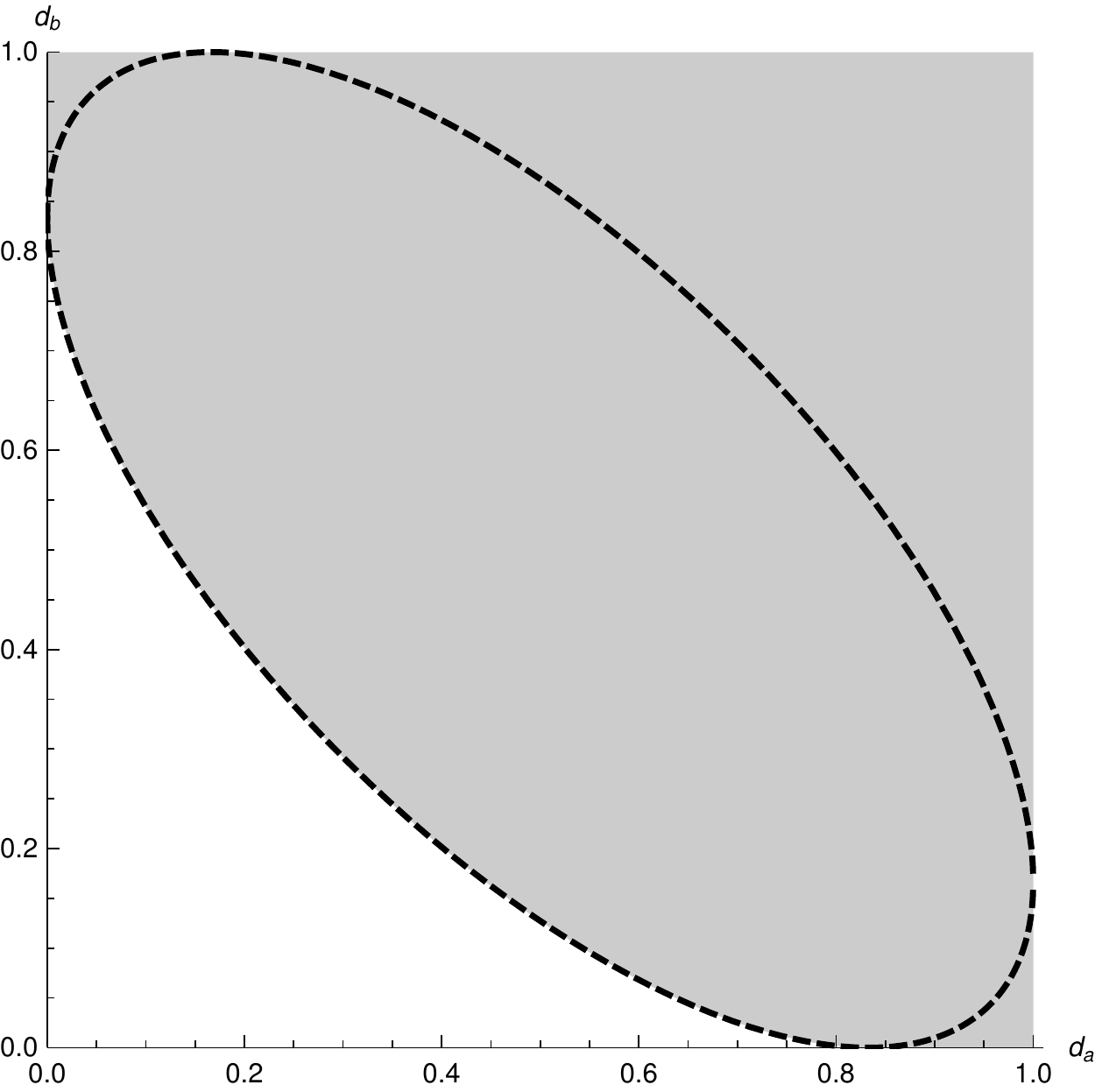}
  \end{subfigure}\\
  \begin{subfigure}[t]{0.4\textwidth}
    \includegraphics[width=\textwidth]{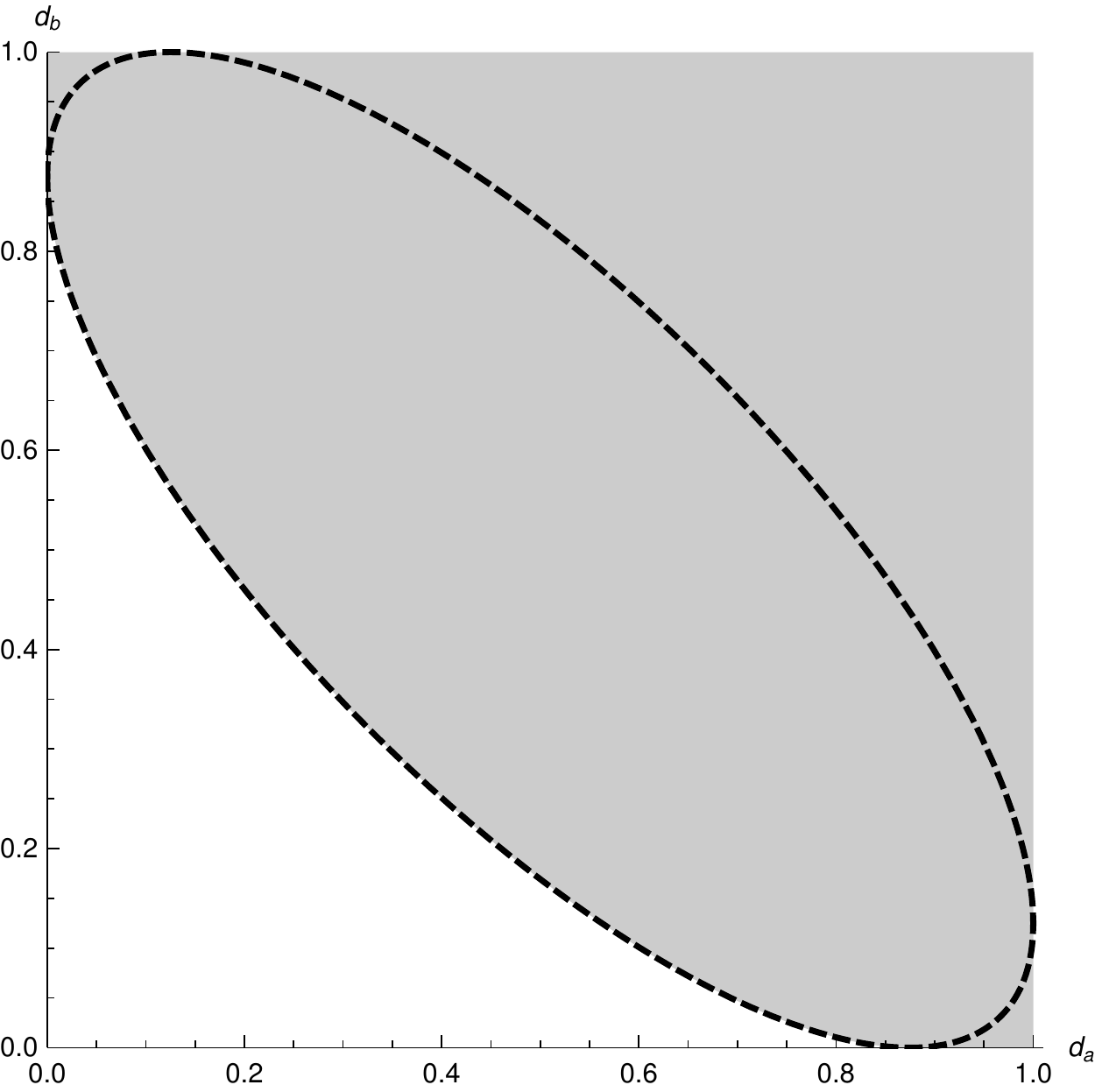}
  \end{subfigure}\quad
  \begin{subfigure}[t]{0.4\textwidth}
    \includegraphics[width=\textwidth]{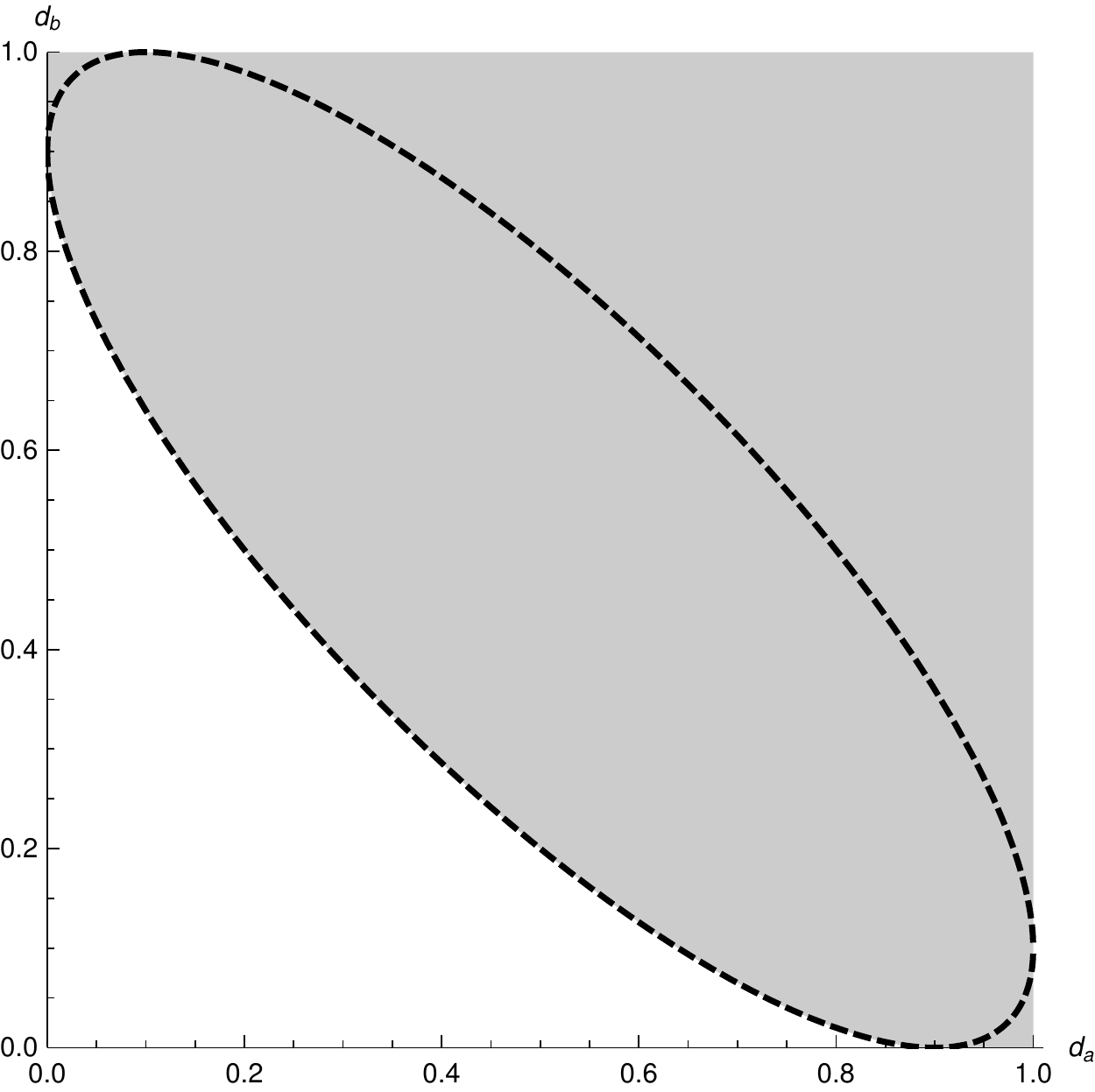}
  \end{subfigure}\\
  \caption{The measurement uncertainty region for quantum Fourier pair observables in several dimensions. The dashed ellipse gives the region explored by covariant observables, whilst the full uncertainty region given by the monotone closure is shaded.}
\end{figure*}

\section{Summary}

We have defined and characterised a covariantisation map, as well as a large set of figures of merit for the approximation of one observable by another which are not increased by the action of the map. This is a useful tool as the space of covariant observables is smaller in dimension than the full set of observables. The covariantisation map may also be used to explore the set of covariant observables, as in \cref{eqn:finite-phase-space-cov-joints}, where the space of covariant joint observables for finite phase space observables is equivalent to the space of density operators. 

The covariantisation map is applicable to both finite and infinite separable Hilbert spaces but is only suitable for observables with finitely many outcomes. Although this set is highly restricted compared to the full set of quantum observables it contains all observables that have thus far been measured in experiment, further, it does not seem likely that an observable with infinitely many outcomes will ever be measured. Even so, it is natural to seek to extend our map to the full set of observables, since in practice one might be interested in  measuring finite outcome observables which are approximations to ideal observables with infinite outcomes. We have discussed some of the properties and restrictions such an extension must have, in particular since the Haar measure of a non-compact group may not be normalised.

Finally we applied this framework to derive measurement uncertainty regions for the three Pauli observables for a qubit Hilbert space, and for the phase space observables of arbitrary finite dimensional phase spaces. The former, simpler example seems to be novel, whilst the latter example has been examined already in~\cite{Werner2016}. Although the underlying definition of the error measure used is different to ours, the two end up being numerically equal in this case. We expect that the framework we have defined may be applied to other problems in measurement uncertainty, and hope that they may be applied more broadly. In particular the covariantisation map may be applicable to many other areas where covariant observables play a role.

\section*{Acknowledgements}
The origin of this manuscript was a series of discussions between the author and Paul Busch. The author acknowledges financial support from EPSRC and the Department of Mathematics of the University of York.

\bibliographystyle{plainnat}
\bibliography{cov-mur}

\appendix

\section{Increasing the error}
\label{sec:inc-error}
Applying the techniques in \cref{sec:invariant-mean} results in compatible approximations that are ``not worse than'' any other compatible approximations, in the sense that for any family of compatible approximations to the targets, there exists a covariant family of compatible approximations with $d_p$ values less than or equal to the original family. It is therefore useful to know when we can \emph{increase} the $d_p$ values so we can cover the entire uncertainty region with convex combinations of covariant and trivial observables.

\begin{lem}[Increasing the error - $\infty$-norm]
  \label{lem:infty-increase-error}
  Let $\{\ope_i\}$ be a family of observables with outcome sets $\Omega_i$. Choose $i\in 1\hdots n$ and $\bov = (v_1\hdots v_i\hdots v_n) \in S_\infty(\ope_1\hdots\ope_n)$, such that there exists some $\omega^*\in\Omega_i$ where $\ope_i(\omega^*)$ is not of full rank, then $v_i \leq v_i^\prime \leq 1 \implies \bov^\prime = (v_1\hdots v_i^\prime\hdots v_n) \in S_\infty(\ope_1\hdots\ope_n)$.
  \begin{proof}
    Let $\Omega = \prod_i \Omega_i$ be the Cartesian product of the outcome sets, since $\bov\in S_\infty(\ope_1\hdots\ope_n)$ there exists a compatible family of observabless $\opf_i$ with joint $\opj:\Omega\to\saops[\hcal]$ such that
    \begin{align}
      d_\infty(\ope_i, \opf_i) = v_i
    \end{align}
    Now define
    \begin{align}
      \tilde{\opj}&:\Omega\to \saops[\hcal]\\
      \tilde{\opj}&:(\omega_1\hdots\omega_{n}) \mapsto \begin{cases}\sum_{\omega\in\Omega_i} \opj(\omega_1\hdots\omega_{i-1},\omega,\omega_{i+1}\hdots\omega_{n}), & \omega_i = \omega^*\\\ 0,&\text{else}\end{cases},
    \end{align}
    Let $\tilde{\opf}_j$ be the $j^\text{th}$ Cartesian margin of $\tilde{\opj}$, and note that for $j\neq i$ we have $\tilde{\opf}_j = \opf_j$, but that $\tilde{\opf}_i:\omega\mapsto \delta_{\omega \omega^*}\opi$ is the trivial observable which gives outcome $\omega^*$ with certainty in any state. Since $\ope_i(\omega^*)$ is not of full rank, there exists a pure state $\rho$ such that $\tr{\ope_i(\omega^*)\rho} = 0$; therefore $d_\infty(\ope_i, \tilde{\opf}_i) = 1$.

    We can now define the observable $\opj_\lambda = (1-\lambda) \opj + \lambda \tilde{\opj}$, for $\lambda\in[0,1]$, with margins $\opf_{j \lambda}$. As before we have $j\neq i \implies \opf_{j\lambda} = \opf_j$, but $\opf_{i,\lambda} = (1-\lambda)\opf_i + \lambda \tilde{\opf}_i$. We can compute the distance
    \begin{align}
      d_\infty(\ope_i, \opf_{i\lambda}) &= \sup_{\rho\in\dops[\hcal]}\max_{\omega\in\Omega} \abs{\tr{\rho(\ope_i(\omega) - \opf_{i\lambda}(\omega))}}\\
                                        &= \sup_{\rho\in\dops[\hcal]}\max_{\omega\in\Omega} \abs{(1-\lambda)\tr{\rho(\ope_i(\omega) - \opf_{i})} + \lambda\tr{\rho(\ope_i(\omega) - \tilde{\opf}_{i})}(\omega)},
    \end{align}
    as we take the $\sup_\rho$ over a compact set, so $\lambda\mapsto d_\infty(\ope_i,\opf_{i\lambda})$ is a continuous function from $[0,1]\to\mathbb{R}^+$, by the intermediate value theorem every value between $d_\infty(\ope_i, \opf_{i})$ and $d_\infty(\ope_i, \tilde{\opf}_i) = 1$ is achieved by some $\lambda$.
  \end{proof}
\end{lem}

\begin{lem}[Increasing the error - $p$-norm]
  \label{lem:p-increase-error}
  Let $\{\ope_i\}$ be a family of observables with outcome sets $\Omega_i$. Choose $i\in 1\hdots n$ and $\bov = (v_1\hdots v_i\hdots v_n) \in S_p(\ope_1\hdots\ope_n)$, such that there exists some $\omega^*\in\Omega_i$ where $\tr{\ope_i(\omega^*)\rho^*} = 1$ for some $\rho^*\in\dops[\hcal]$ then $v_i \leq v_i^\prime \leq 2^{\frac{1}{p}} \implies \bov^\prime = (v_1\hdots v_i^\prime\hdots v_n) \in S_p(\ope_1\hdots\ope_n)$.
  \begin{proof}
    Let $\Omega = \prod_i \Omega_i$ be the Cartesian product of the outcome sets, since $\bov\in S_p(\ope_1\hdots\ope_n)$ there exists a compatible family of observables $\opf_i$ with joint $\opj:\Omega\to\saops[\hcal]$ such that
    \begin{align}
      d_p(\ope_i, \opf_i) = v_i
    \end{align}
    Now choose $\tilde{\omega} \neq \omega^*$ and define
    \begin{align}
      \tilde{\opj}&:\Omega\to \saops[\hcal]\\
      \tilde{\opj}&:(\omega_1\hdots\omega_{n}) \mapsto \begin{cases}\sum_{\omega\in\Omega_i} \opj(\omega_1\hdots\omega_{i-1},\omega,\omega_{i+1}\hdots\omega_{n}), & \omega_i = \tilde{\omega}\\\ 0,&\text{else}\end{cases},
    \end{align}
    Let $\tilde{\opf}_j$ be the $j^\text{th}$ Cartesian margin of $\tilde{\opj}$, and note that for $j\neq i$ we have $\tilde{\opf}_j = \opf_j$, but that $\tilde{\opf}_i:\omega\mapsto \delta_{\omega \tilde{\omega}}\opi$ is the trivial observable which gives outcome $\tilde{\omega}$ with certainty in any state. Since we have $\tr{\ope(\omega^*)\rho^*} = 1$ we can compute 
    \begin{align}
      d_p(\ope_i,\tilde{\opf}_i) &= \sup_\rho\left(\sum_{\omega\in\Omega_i}\abs{\tr{\rho(\ope_i(\omega)- \tilde{\opf}_i(\omega))}}^p\right)^\frac{1}{p}\\
                                 &\geq \left(\sum_{\omega\in\Omega_i}\abs{\tr{\rho^*(\ope_i(\omega)- \tilde{\opf}_i(\omega))}}^p\right)^\frac{1}{p}\\
                                 &= \left(\abs{\tr{\rho^*\tilde{\opf}_i(\tilde{\omega})}}^p + \abs{\tr{\rho^*\ope_i(\omega^*)}}^p\right)^\frac{1}{p}\\
                                 &= 2^\frac{1}{p}
    \end{align}

    We can now define the observable $\opj_\lambda = (1-\lambda) \opj + \lambda \tilde{\opj}$, for $\lambda\in[0,1]$, with margins $F_{j \lambda}$. As before we have $j\neq i \implies \opf_{j\lambda} = \opf_j$, but $\opf_{i,\lambda} = (1-\lambda)\opf_i + \lambda \tilde{\opf}_i$. We can compute the distance
    \begin{align}
      d_p(\ope_i, \opf_{i,\lambda}) &= \sup_\rho\left(\sum_{\omega\in\Omega_i}\abs{\tr{\rho(\ope_i(\omega)- \tilde{\opf}_{i\lambda}(\omega))}}^p\right)^\frac{1}{p},
    \end{align}
    as we take the $\sup_\rho$ over a compact set, $\lambda\mapsto d(\ope_i,\opf_{i,\lambda})$ is a continuous function from $[0,1]\to\mathbb{R}^+$. By the intermediate value theorem every value between $d(\ope_i, \opf_{i})$ and $d(\ope_i, \tilde{\opf}_i) = 2^\frac{1}{p}$ is achieved by some $\lambda$.
  \end{proof}
\end{lem}

\section{Computing the uncertainty region for finite space space observables}
\label{sec:app:computing-fourier-ur}

\subsection{Commutivity} 
There is a one-to-one relation between covariant joint observables $\opj:\cyc[n]\times\cyc[n]\to\posops[\hcal]$ and trace one positive operators on $\hcal$ given by 
\begin{align}\label{eqn:finite-phase-space-cov-joints}
  \opj:(k,q)\mapsto \frac{1}{n}R_{k,q}\left[\tau\right].
\end{align}
All covariant, $\cyc[n]\times\cyc[n]$ valued observable are obtained in this way, for some trace $1$ positive $\tau$, as we can take $\tau = n\opj(0,0)$, and all trace $1$ positive operators give rise to some covariant, $\cyc[n]\times\cyc[n]$ valued observable. We can write down the margins of such an observable
\begin{align}
  \opc&:\cyc[n] \to \posops[\hcal] & \opd&:\cyc[n] \to \posops[\hcal]\\
  \opc&:g\mapsto \sum_h \opj(g,h) = \frac{1}{n}\sum_h R_{g,h}\left[\tau\right] &  \opd&:h\mapsto \sum_g \opj(g,h) = \frac{1}{n}\sum_g R_{g,h}\left[\tau\right].
  \label{eqn:write-down-margins}
\end{align}
We can show that each $\opc(k)$ commutes with each $V_q$
\begin{align}
  \opc(g) &= \sum_h \opj(g,h)\\
          &= \sum_h \opj(g, h+q)\\
          &= \sum_h R_{0,q}[\opj(g,h)]\\
          &= V_q \opc(g) V_q^*\\
  \implies \opc(g) V_q &= V_q \opc(g), \quad \forall g,q\in \cyc[n].
\end{align}
A similar calculation gives
\begin{align}
  \opd(h) U_k = U_k \opd(h), \quad \forall h,k\in \cyc[n].
\end{align}
Indeed an explicit calculation gives

\begin{align}
  \opc(g) &= \sum_{k} \ketbra{k+g}{k+g} \bra{k} \tau \ket{k}\\
  \opd(h) &= \sum_{q} \ketbra{f_{q+h}}{f_{q+h}} \bra{f_{q}} \tau \ket{f_{q}}.
\end{align}
\subsection{Computing the sup-norm}
The simultaneous diagonalisability of $\opa$ and $\opc$ allows us to compute $d_\infty(\opa,\opc)$ explicitly. Without loss of generality let
\begin{align}
  \opc(0) = \sum_{k\in\cyc[n]} c_k \ketbra{k}{k}
\end{align}
for $c_k\in \left[0,1\right]$, and $\sum_k c_k = 1$. Then
\begin{align}
  d_\infty(\opa,\opc) &= \sup_\rho \max_{g} \abs{\tr{\rho\left[\opa(g) - \opc(g)\right]}} \\
                      &=  \sup_\rho \max_g \abs{\tr{\rho\left[\ketbra{g}{g} - U_g\sum_{k\in\cyc[n]}c_k\ketbra{k}{k} U_g^\dagger\right]}} \\
                      &=  \sup_\rho \max_g \abs{\tr{\rho U_g\left[\ketbra{0}{0} - \sum_{k\in\cyc[n]}c_k\ketbra{k}{k} \right]U_g^\dagger}} \\
                      &=  \sup_\rho \max_g \abs{\tr{U_g^\dagger \rho U_g\left[\ketbra{0}{0} - \sum_{k\in\cyc[n]}c_k\ketbra{k}{k} \right]}} \\
                      &=  \sup_\rho \abs{\tr{\rho \left[\ketbra{0}{0} - \sum_{k\in\cyc[n]}c_k\ketbra{k}{k} \right]}} \\
                      &=  \max\{1-c_0, c_1, \hdots, c_{n-1}\}.
\end{align}
Now note that
\begin{align}
  \sum_{k\in\cyc[n]} c_k = 1 \implies \sum_{k\neq 0} c_k = 1-c_0
\end{align}
combined with $c_k \geq 0$ we see that
\begin{align}
  1-c_0 \geq c_k, \quad \forall k > 0,
\end{align}
so
\begin{align}
  \label{eqn:comp-constraint}
  d_\infty(\opa,\opc) = 1-c_0.
\end{align}
Similarly, if
\begin{align}
  \opd(0) = \sum_{r\in\cyc[n]} d_r \ketbra{f_r}{f_r},
\end{align}
then
\begin{align}
  \label{eqn:four-constraint}
  d_\infty(\opb,\opd) = 1-d_0.
\end{align}
\subsection{Semidefinite program}
We can use relations \eqref{eqn:comp-constraint} and \eqref{eqn:four-constraint}, along with \eqref{eqn:write-down-margins} to put constraints on the operator $\tau$ we used to define the joint
\begin{align}
  \sum_h \opj(g,h) &= \opc(g) = U_g \opc(0) U_g^\dagger\\
  \frac{1}{n} \sum_h U_gV_h \tau V_h^\dagger U_g^\dagger &= U_g \opc(0) U_g^\dagger\\
  \iff \frac{1}{n} \sum_h V_h \tau V_h^\dagger  &=  \opc(0)  \\
  \frac{1}{n} \sum_g U_g \tau U_g^\dagger  &=  \opd(0).
\end{align}
Computing matrix elements gives
\begin{align}
  \bra{k} \opc(0) \ket{l} = \frac{1}{n} \sum_h \bra{k} V_h \tau V_h^\dagger \ket{l} &= \frac{1}{n} \sum_h \bra{k} V_h^\dagger \tau V_h \ket{l} \\
                                                                          &= \frac{1}{n} \sum_h \bra{k}\tau \ket{l} e^{\frac{2\pi i}{n} h(l-k)}\\
                                                                          &= \bra{k}\tau \ket{l} \delta_{k,l}\\
  \bra{f_r} \opd(0) \ket{f_s} &= \bra{f_r}\tau \ket{f_s} \delta_{r,s}.
\end{align}
Given that the only matrix elements that affect the uncertainties are the $(0,0)$ matrix element of $\opc(0)$ and the $(f_0, f_0)$ matrix element of $\opd(0)$ the relevant constraints are
\begin{align}
  \bra{0}\tau\ket{0} &= 1 - d_\infty(\opa,\opc)\\
  \sum_{k,l} \bra{k}\tau\ket{l} &=  n(1- d_\infty(\opb,\opd)).
\end{align}
If we set 
\begin{align}
  A_n &= \sum_{k,l} \ketbra{k}{l}
\end{align}
then computing the lower boundary of the uncertainty region is equivalent to the following semidefinite program, for each $d_a\in [0,1]$
\begin{equation}
  \begin{aligned}
    & \underset{X}{\text{maximise}}
    & & p = \tr{A_n X} \\
    & \text{subject to}
    & \tr{\ketbra{0}{0} X} &= 1-d_a, \\
    &&  \tr{\opi_n  X} &= 1, \\
    && X &\geq 0.
  \end{aligned}
  \label{eqn:semidefinite-program}
\end{equation}
We can impose the equality constraints in \eqref{eqn:semidefinite-program}, by means of the linear map
\begin{align}
  \mathcal{M}&: \saops[\hcal] \to M_2(\mathbb{C})\\
  \mathcal{M}&:X \mapsto \begin{pmatrix}\tr{\ketbra{0}{0} X} & 0 \\ 0 & \tr{X}\end{pmatrix},
\end{align}
where $M_2(\mathbb{C})$ is the set of $2$ by $2$ matrices over the field $\mathbb{C}$. If
\begin{align}
  B = \begin{pmatrix}1-d_a & 0 \\ 0 & 1\end{pmatrix}
\end{align}
then the equality constraints are
\begin{align}
  \mathcal{M}(X) = B
\end{align}
We can compute the dual of $\mathcal{M}$ directly from the defining relation
\begin{align}
  \tr{\mathcal{M}^*(Y) X} &= \tr{Y \mathcal{M}(X)}\\
                          &= Y_{00} \tr{\ketbra{0}{0}X} + Y_{11} \tr{\opi_n X} \\
  \mathcal{M}^*\left(\begin{pmatrix}Y_{00} & Y_{01} \\ Y_{10} & Y_{11}\end{pmatrix}\right) &= Y_{00} \ketbra{0}{0} + Y_{11}\opi_n.
\end{align}
The dual problem to \eqref{eqn:semidefinite-program} is then given by
\begin{equation}
  \begin{aligned}
    & \underset{Y}{\text{minimise}}
    & d &= \tr{B Y} \\
    & \text{subject to}
    & \mathcal{M}^*(Y) &\geq A_n\\
    && Y &\in M_2(\mathbb{C}).
  \end{aligned}
  \label{eqn:semidefinite-program-dual}
\end{equation}
Alternatively 
\begin{equation}
  \begin{aligned}
    & \underset{y_0, y_1\in\mathbb{R}}{\text{minimise}}
    & d &= (1-d_a)y_0 + y_1 \\
    & \text{subject to}
    & 0&\leq y_0 \ketbra{0}{0} + y_1 \sum_k \ketbra{k}{k} - \sum_{k,l} \ketbra{k}{l} = Z.
  \end{aligned}
\end{equation}
It is easy to see that we have \emph{strong duality} for these problems, since we can always choose $y_1$ large enough that $Z > 0$, by the Slater condition~\cite{rtr-conv-anal-book} we therefore know that wherever the solution $d$ to the dual problem is finite we have that $\inf d=\sup p$.


Henceforth we mix operators interchangeably with their matrices in the computational basis. Define the \emph{characteristic polynomial} function for each $n\in\mathbb{N}$
\begin{align}
  \chi_n:M_n(\mathbb{C})\times \mathbb{R} &\to \mathbb{R}\\
  \chi_n(X, x) &= \det(x\opi_n - X).
\end{align}
We can compute the characteristic polynomial of the matrix $Z$
\begin{align}
  \chi_n(Z,x) &= \det\left(x\opi_n -Z\right)\\
              &=\det\left((x-y_1)\opi_n - y_0\ketbra{0}{0} + A_n\right)\\
              &=\det\left((x-y_1)\opi_n + A_n\right) -y_0 \bra{0}\adj{(x-y_1)\opi_n + A_n} \ket{0}\\
              &=\det\left((x-y_1)\opi_n + A_n\right) -y_0 \det\left((x-y_1)\opi_{n-1} + A_{n-1}\right)\\
              &=(-1)^n \det\left((y_1-x)\opi_n - A_n\right) -(-1)^{n-1} y_0 \det\left((y_1-x)\opi_{n-1} - A_{n-1}\right)\\
              &=(-1)^n \chi_n(A_n,y_1-x)  + (-1)^{n} y_0 \chi_{n-1}(A_{n-1}, y_1-x)\\
              &= (-1)^n\left[(x-y_1-n)(x-y_1)^{n-1} +y_0(x-y_1-n+1)(x-y_1)^{n-2}\right]\\
              &= (-1)^n(x-y_1)^{n-2}\left[(x-y_1-n)(x-y_1) -y_0(x-y_1-n+1)\right]\\
              &= (-1)^n(x-y_1)^{n-2}\left[x^2 + x(n-y_0 -2y_1) + \left(y_1^2 + y_1(y_0-n) + y_0(1-n)\right)\right]
                \label{eqn:char-poly}
\end{align}
where $\operatorname{adj}$ denotes the adjudicate matrix, and we have employed the classical matrix determinant lemma, as well as the fact that
\begin{align}
  \chi_n(A_n,x) = (x-n)x^{n-1},
\end{align}
for $A_n$ the $n$ by $n$ matrix of ones~\cite{matrix-analysis}.
We are seeking constraints on $y_0$ and $y_1$ which are necessary and sufficient for all of the roots of $x\mapsto \chi_n(Z,x)$ to be non-negative, we can read off from \eqref{eqn:char-poly} that $y_1 \geq 0$. We now need to examine the roots of 
\begin{align}
  x\mapsto x^2 + x(n-y_0 -2y_1) + \left(y_1^2 + y_1(y_0-n) + y_0(1-n)\right),
\end{align}
the quadratic formula gives
\begin{align}
  x^\pm = \frac{1}{2}\left(y_0 +2y_1 -n \pm \sqrt{(y_0 + 2y_1 -n)^2 - 4(y_1^2 + y_1(y_0-n) + y_0(1-n))} \right),
\end{align}
note that the roots are automatically real, as our matrices are self-adjoint. The $x^\pm$ are both non-negative if, and only if
\begin{align}
  \sqrt{(y_0 + 2y_1 -n)^2 - 4(y_1^2 + y_1(y_0-n) + y_0(1-n))} \leq y_0 +2y_1 -n,
\end{align}
which is satisfied if and only if
\begin{align}
  0&\leq y_0 +2y_1 -n,
     \label{eqn:y1-consraint-1}
\end{align}
and
\begin{align}
 0&\leq y_1^2 + y_1(y_0-n) + y_0(1-n),
    \label{eqn:y1-constraint-2}
\end{align}
are both satisfied. The solutions of
\begin{align}
  y_1^2 + y_1(y_0-n) + y_0(1-n) = 0
\end{align}
are
\begin{align}
  y_1^\pm = \frac{1}{2}\left(n-y_0 \pm \sqrt{(n-y_0)^2 - 4y_0(1-n)}\right).
\end{align}
It is easy to show that the radicant is positive. The constraint in \eqref{eqn:y1-constraint-2} is therefore satisfied if, and only if
\begin{align}
  y_1 \geq \frac{1}{2}\left(n-y_0 + \sqrt{(n-y_0)^2 + 4y_0(n-1)}\right)
\end{align}
or
\begin{align}
  y_1 \leq \frac{1}{2}\left(n-y_0 - \sqrt{(n-y_0)^2 + 4y_0(n-1)}\right)
\end{align}
Rewriting \eqref{eqn:y1-consraint-1} we see we need
\begin{align}
  y_1 \geq \frac{1}{2}\left(n-y_0\right),
\end{align}
therefore all of the constraints are satisfied if, and only if
\begin{align}
  y_1 \geq \frac{1}{2}\left(n-y_0 + \sqrt{(n-y_0)^2 + 4y_0(n-1)}\right),
\end{align}
since the quantity on the right hand side is always positive. Recall that we are attempting to minimise the quantity
\begin{align}
  d &= (1-d_a)y_0 + y_1,
\end{align}
subject to the positivity constraints. We therefore choose
\begin{align}
  y_1 &= \frac{1}{2}\left(n-y_0 + \sqrt{(n-y_0)^2 + 4y_0(n-1)}\right)\\
  \implies d &= \left(\frac{1}{2} - d_a\right)y_0 + \frac{1}{2}\left(n+\sqrt{(n-y_0)^2 + 4y_0(n-1)}\right),
\end{align}
differentiating, we find that $d$ is minimised where
\begin{align}
  y_0 = 2-n - \abs{1-2d_a}\sqrt{\frac{n-1}{d_a(1-d_a)}},
\end{align}
and that at this point 
\begin{align}
  d &= 1 + d_a(n-2) + 2\sqrt{d_a(1-d_a)(n-1)}\\
  \implies d_b^{\text{min}} &= 1- \frac{d}{n}\\
    &= 1 -  \frac{1}{n}\left(1 + d_a(n-2) + 2\sqrt{d_a(1-d_a)(n-1)}\right).
\end{align}
We note that this is a section of the ellipse with defining equation
\begin{align}
0 = n^2d_a^2 + n^2 d_b^2 + 2n(n-2)d_ad_b + 2n(1-n)d_a + 2n(1-n)d_b  + (n-1)^2,
  \label{eqn:ellipse-defining-equation}
\end{align}
which has center $\left(\frac{1}{2}, \frac{1}{2}\right)$, and touches the coordinate axes at the points $\left(0, 1-\frac{1}{n}\right)$ and $\left(1-\frac{1}{n}, 0\right)$. The major axis of the ellipse has angle $\frac{\pi}{4}$ with each coordinate axis, as it must by symmetry.

\end{document}